\documentclass[%
superscriptaddress,
preprint,
 amsmath,amssymb,
 aps,
prl,
]{revtex4-1}

\usepackage{graphicx}
\usepackage{dcolumn}
\usepackage{bm}
\usepackage{xr}

\makeatletter
\newcommand*{\addFileDependency}[1]{
  \typeout{(#1)}
  \@addtofilelist{#1}
  \IfFileExists{#1}{}{\typeout{No file #1.}}
}
\makeatother

\newcommand*{\myexternaldocument}[1]{%
    \externaldocument{#1}%
    \addFileDependency{#1.tex}%
    \addFileDependency{#1.aux}%
}
\myexternaldocument{methods}

\begin{document}

\title{Charge order textures induced by non-linear lattice coupling in a half-doped manganite}

\author
{Ismail El Baggari,$^{1,\ast}$ 
David J. Baek,$^{2,a,\ast}$ 
Michael J. Zachman,$^{3,b}$ 
Di Lu,$^{4,5,c}$ 
Yasuyuki Hikita,$^{5,6}$ 
Harold Y. Hwang,$^{5,6}$ 
Elizabeth A. Nowadnick,$^{7}$ 
Lena F. Kourkoutis$^{3,8}$\\ 
\phantom{breakline}
\newline
\textit{$^{1}$Department of Physics, Cornell University, Ithaca, NY}\\
\textit{$^{2}$School of Electrical and Computer Engineering, Cornell University, Ithaca, NY}\\
\textit{$^{3}$School of Applied and Engineering Physics, Cornell University, Ithaca, NY}\\
\textit{$^{4}$Department of Physics, Stanford University, Stanford, CA}\\
\textit{$^{5}$Stanford Institute for Materials and Energy Sciences, SLAC National Accelerator Laboratory, Menlo Park, CA}\\
\textit{$^{6}$Department of Applied Physics, Stanford University, Stanford, CA}\\
\textit{$^{7}$Department of Materials Science and Engineering, University of California Merced, Merced CA}\\
\textit{$^{8}$Kavli Institute at Cornell for Nanoscale Science, Cornell University, Ithaca, NY}\\
\phantom{breakline}
\phantom{breakline}\\
\small{$^{a}$ Current address: Intel Corp., Hillsboro, OR}\\
\small{$^{b}$ Current address: Center for Nanophase Materials Sciences, Oak Ridge National Laboratory, TN}\\
\small{$^{c}$ Current address: Materials Science and Engineering, Northwestern University, Evanston, IL}\\
\small{$^\ast$ These authors contributed equally to this work.}\\
}

\maketitle

\textbf{
The self-organization of strongly interacting electrons into superlattice structures underlies the properties of many quantum materials. 
How these electrons arrange within the superlattice dictates what symmetries are broken and what ground states are stabilized. 
Here we show that cryogenic scanning transmission electron microscopy enables direct mapping of local symmetries and order at the intra-unit-cell level in the model charge-ordered system Nd$_{1/2}$Sr$_{1/2}$MnO$_{3}$.
In addition to imaging the prototypical site-centered charge order, we discover the nanoscale coexistence of an exotic intermediate state which mixes site and bond order and breaks inversion symmetry. 
We further show that nonlinear coupling of distinct lattice modes controls the selection between competing ground states.
The results demonstrate the importance of lattice coupling for understanding and manipulating the character of electronic self-organization and highlight a novel method for probing local order in a broad range of strongly correlated systems. 
}
\clearpage

\begin{figure}
  \includegraphics[width=.7\linewidth]{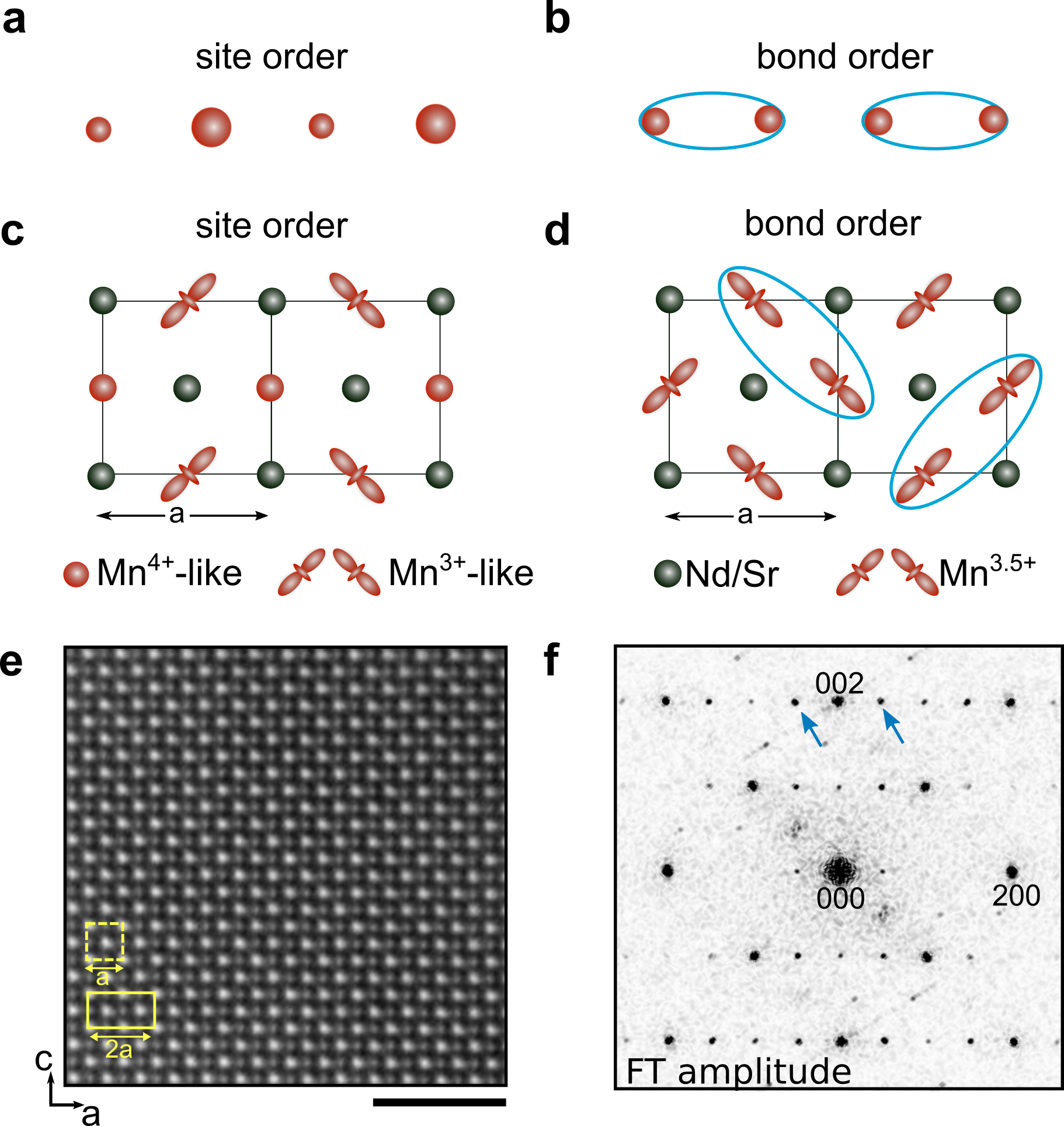}
  \caption{
  \textbf{Charge order supercell}.
   \textbf{a}, Charge order modulation centered on sites (red circles).
   The size of the circle indicates the electronic density on the site.
  \textbf{b}, Charge order modulation centered on the bonds. 
  Ellipses represent stronger, more electron-rich bonds.
  The sites remain equivalent in density.
  \textbf{c}, Prototypical site-centered charge and orbital order model in the half-doped manganites, including orbital order.
  \textbf{d}, Alternative bond-centered model in which all Mn sites have the same charge configuration (Mn$^{3.5+}$).
  In this case, the supercell is generated by the orbital order.
  \textbf{e}, High-angle annular dark-field (HAADF) scanning transmission electron microscopy (STEM) projection image at 93 K, where we expect the system to be in the charge-ordered phase.
  Bright atomic columns represent Nd/Sr sites and dark atomic columns represent Mn sites. 
  The scale bar corresponds to 2 nm.
  \textbf{f}, Fourier transform amplitude of the STEM image. 
  At 93 K, superlattice peaks (arrows) with $\mathbf{Q}^{CO}$ = (1/2, 0, 0) are evident, indicating the formation of a twofold superlattice in real space (box in E).
  }
  \label{F:Fig1}
\end{figure}

Strong interactions between electrons and the atomic lattice often lead to their self-organization into ordered spatial patterns \cite{tranquada1995evidence,mori1998pairing,abbamonte2004crystallization,sasaki2017crystallization}. 
One well-known example is charge ordering, the spatial modulation of the electronic charge density which forms superlattices and governs the properties of many exotic materials, from oxides to transition-metal chalcogenides to charge-transfer salts \cite{chang2012direct,li2016controlling,gao2013mapping}. 
In general, charge ordering is studied at scales larger than the superlattice, with a focus on the average periodicity of the charge modulation or the degree of long range order \cite{milward2005electronically,comin2015broken}. 
On the other hand, the microscopic arrangement at sub-unit-cell length scales, such as whether the electrons reside on the atomic sites or bonds (Fig.~\ref{F:Fig1}a and b), is significantly more challenging to measure, but dictates the symmetry of the system in addition to the mechanism underlying electronic order \cite{Comin2015,Achkar2016,zhao2019charge}.

Perovskite $3d$ transition-metal oxides are a class of materials in which charge order plays an especially important role, influencing antiferromagnetic order, high-temperature superconductivity, and colossal magnetoresistance \cite{tranquada1994simultaneous,chang2012direct,Uehara1999}. 
The strong hybridization between the transition metal (site) and oxygen (bond) in these systems contributes to their rich, intricate electronic structure landscape \cite{khomskii2014transition}. 
In cuprates, for instance, recent bulk measurements suggest that the charge-ordered phase may contain a site-centered modulation in addition to the putative bond-centered $d$-wave modulation \cite{Comin2015,Mcmahon2019}. 
A similarly complex and fundamental debate concerns the half-doped manganites, an ideal playground for exploring coupled charge and orbital orders and phase competition \cite{tokura2000orbital}.
The widely accepted model, proposed over 60 years ago, invokes a site-centered modulation in which electrons localize on manganese sites and form a zig-zag orbital order pattern which doubles the unit cell (Fig.~\ref{F:Fig1}c) \cite{Goodenough1955,Radaelli1997,goff2004charge}. 
An alternative bond-centered state has also been proposed \cite{Daoud2002}; 
in this case a charge modulation does not occur so the superlattice is generated from orbital ordering (Fig.~\ref{F:Fig1}d).
Subsequent experimental studies, however, have not conclusively confirmed this scenario \cite{goff2004charge,Rodriguez2005,jooss2007polaron}. 
Theory further predicts an even more exotic intermediate state which mixes both site and bond characters, leading to the breaking of inversion symmetry through the formation of uncompensated electric dipoles \cite{Efremov2004,VanDenBrink2008}. 
Determining if such a state exists or whether charge order corresponds to a pure site- or bond-centered state remains a fundamental challenge, with broad and urgent implications for other classes of materials.   

Manganite compounds exhibit large interactions between the lattice and the electronic degrees of freedom, and so most experimental proposals for site- or bond-centered charge order models have relied on obtaining the average lattice distortions and crystal symmetry \cite{Radaelli1997,Daoud2002}. 
Here we employ a novel methodology for probing charge order, based on atomic-resolution cryogenic scanning transmission electron microscopy (cryo-STEM) and the direct measurement of picoscale lattice displacements in real space. 
By mapping the displacement patterns within the charge order superlattice, this method enables us to visualize distinct, coexisting arrangements of charge order and, importantly, establish the existence of the intermediate phase.
Absent the real space visualization, this phase would remain hidden, partly due to the challenge of disentangling nanoscale domains of pure bond/site order from genuine mixed states in bulk-averaged measurements. 
Going further, cryogenic STEM imaging and theoretical calculations reveal coupled secondary order parameters that govern the nature of the charge order phase. 
These atomic-scale visualizations not only advance our understanding of charge ordering through a novel lens but also reveal a tuning knob for manipulating electronic self-organization via lattice coupling.

The material we focus on is a Nd$_{1/2}$Sr$_{1/2}$MnO$_{3}$ thin film which was grown using pulsed laser deposition on a (110)-oriented SrTiO$_{3}$ substrate. 
The particular substrate orientation, which imparts in-plane anisotropy on the film, reproduces the electronic and structural transitions found in the bulk counterpart including the charge order transition~\cite{ogimoto2005strain,lai2010mesoscopic,Kuwahara1995}. 
The temperature dependence of the magnetization in this film matches that of previous reported epitaxial films.
Figure~\ref{F:Fig1}e shows a projection HAADF-STEM image of the crystal along the [010] orientation (in the $Pnma$ space group setting) below the charge order transition temperature.
The atomically-resolved Nd/Sr columns appear brighter than the Mn columns because the contrast in HAADF scales strongly with the atomic number.
Near the crystalline peaks in the Fourier transform amplitude (Fig.~\ref{F:Fig1}f), superlattice peaks (blue arrows) appear at low temperature, indicating the formation of a modulated structure.
These are located at $\mathbf{Q}^{CO}$ = $(1/2,0,0)$ as expected for charge ordering at half doping.
The modulation is unidirectional, however, some regions of the sample exhibit bi-directional modulations.
These might arise from the coexistence of small charge order domains (either in-plane or out-of-plane) within a single crystal twin or from the presence of crystalline twins that establish the direction of the charge order wavevector.
By mapping both the charge order and the crystalline order parameters, we find that the orthogonal charge order domains are coupled to crystalline twins in the sample (Supplementary Fig. 3).

The exact intra-unit-cell arrangement within charge order superlattices is key to understanding their microscopic origins and interactions with other electronic phases.
In the majority of theoretical treatments of charge-ordered phases, the ground state is discussed in terms of pure electronic degrees of freedom.
Site-centered charge order in half-doped manganites, for instance, is described as the alternation of Mn$^{4+}$ and Mn$^{3+}$ species \cite{Goodenough1955,Radaelli1997}; 
however, the degree of charge disproportionation is much smaller and is better described by Mn valences of $3.5 + \delta$ (Mn$^{4+}$-like) and $3.5 - \delta$ (Mn$^{3+}$-like) with $\delta \ll$ 0.5 \cite{goff2004charge,Grenier2004}.
In reality, the crystal structure also undergoes a variety of complex atomic displacements, such as Jahn-Teller and breathing distortions, which alter the bonding network and hence the electronic configuration.
Therefore, emergent charge and orbital textures are closely linked to the pattern and symmetry of said displacements \cite{egami1993lattice,khalyavin2020emergent}.
For site-centered order, bulk X-ray or neutron structural data suggests that the crystal adopts $P2_{1}/m$ space group symmetry with bond distortions consistent with charge localization on the Mn sites \cite{Radaelli1997,Rodriguez2005,goff2004charge}.
The experimental report for the bond-centered model, on the other hand, found a different space group symmetry ($Pnm2_1$) and a distinct displacement pattern associated with it \cite{Daoud2002}.
The ability to probe \textit{intra}-unit-cell lattice distortions can therefore determine or even reveal novel ground states.

\begin{figure}[h]
  \includegraphics[width=0.6\linewidth]{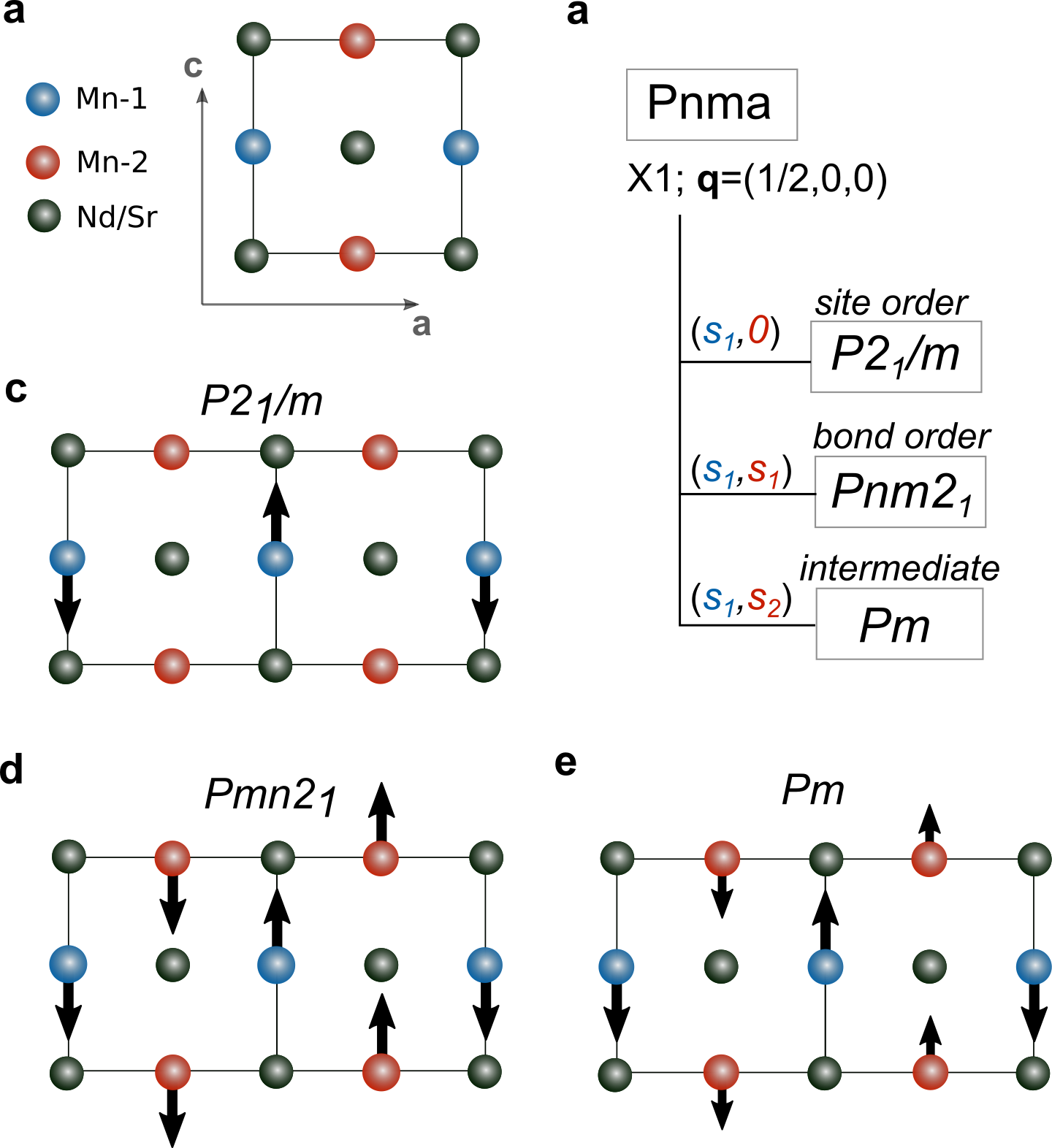}
  \caption{
  \textbf{The symmetry of distorted structures.} 
  \textbf{a}, High-symmetry unit cell with two Mn sublattices.
  \textbf{b}, Group-subgroup relations for displacements with the wavevector $\mathbf{q} = (1/2,0,0)$ and irreducible representation X1.
  The X1 order parameter is two dimensional with each dimension corresponding to a Mn sublattice.
  \textbf{c-e}, Displacement patterns for the site-centered ($s_{1}$,$s_{2}$=$0$), bond-centered ($s_{1}$,$s_{2}=s_{1}$), and intermediate ($s_{1}$,$s_{2}\neq s_{1}$) phase, respectively.
  }
  \label{F:Fig2}
\end{figure}

To firmly connect patterns of atomic displacements to the reported crystal symmetries and hence to the models of electronic order, we first explore the possible distorted structures that emerge from the high-symmetry phase (space group $Pnma$) using group theory.
Any distorted structure must double the unit cell, therefore we require atomic displacements whose wavevector is $\mathbf{q}$ = (1/2, 0, 0).
The relevant distortion consistent with this requirement is the X1 displacement mode, or irreducible representation (irrep), which affects two inequivalent Mn sublattices in the high-symmetry unit cell (See Methods section 4).
In other words, the X1 irrep is two-dimensional with the first dimension corresponding to the first Mn sublattice (Mn-1 in blue) and the second dimension to the second Mn sublattice (Mn-2 in red), as shown in Fig.~\ref{F:Fig2}a.
If X1 displacements, which consist of a complex set of atomic distortions including the transverse Mn displacements shown in Fig.~\ref{F:Fig2}, affect only the first sublattice, the resulting crystal structure has $P2_{1}/m$ symmetry, which is consistent with site-centered order (Fig.~\ref{F:Fig2}b,c).
If they occur in both sublattices and with equal magnitude (Fig.~\ref{F:Fig2}b,d), the resulting crystal structure has $Pnm2_1$ symmetry, which matches that of bond-centered order.
A third structure with $Pm$ symmetry can be obtained by having displacements on both sublattices but with different magnitude (Fig.~\ref{F:Fig2}b,e).
Such an intermediate state combines aspects of both site and bond order, however, experimental refinements of atomic positions in this symmetry have not been performed so far. 
The group theory analysis thus shows that the pattern of Mn displacements uniquely defines the character of the charge-ordered phase, without resorting to measurements of the electronic charge.

We now pinpoint the underlying ordering model in Nd$_{1/2}$Sr$_{1/2}$MnO$_{3}$, by mapping the picoscale lattice degrees of freedom using HAADF-STEM at low temperature.
A reference image lacking the periodic modulation is generated by removing the contribution of the $\mathbf{Q}^{CO}$ superlattice peaks in the Fourier transform \cite{Savitzky2017,ElBaggari2018}.
The displacements are then extracted by mapping the atomic positions in the original image and the reference image. 
Figure~\ref{F:Fig3}a shows a STEM image overlaid with arrows indicating Mn column displacements.
The area of the triangle denotes the magnitude of displacement and the color represents the angle of the displacement relative to the wavevector.
The dominant displacements have transverse polarization and generate a twofold superlattice. 

\begin{figure}
  \includegraphics[width=\linewidth]{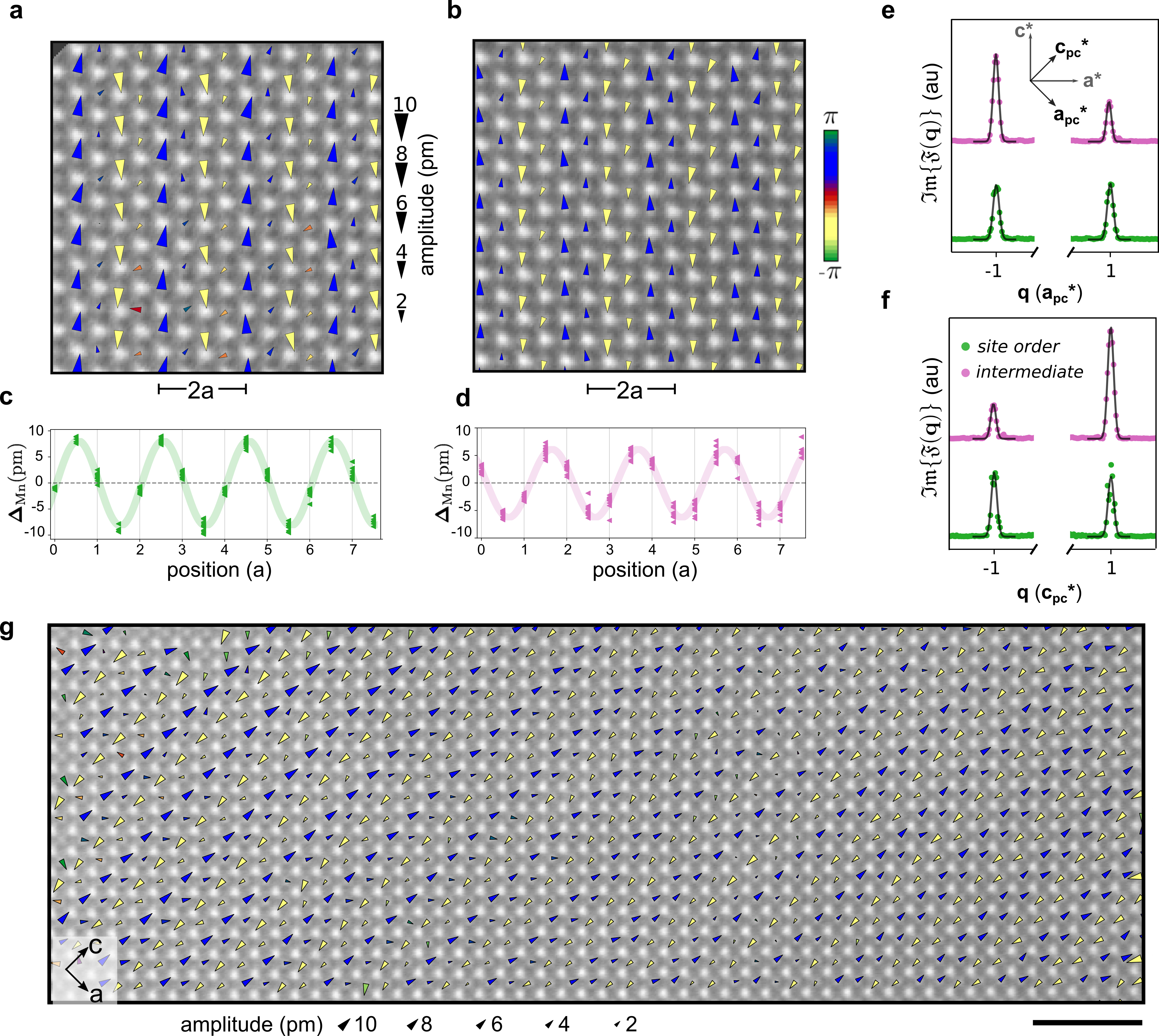}
  \caption{
  \textbf{Nanoscale coexistence of site order and non-centrosymmetric intermediate order.}
  \textbf{a}, Map of periodic lattice displacements of manganese atomic columns ($\mathbf{\Delta}_{\text{Mn}}$).
  The area of the arrows scales with the amplitude of the distortion and the color represents the transverse polarization of the displacements.
   The displacement pattern within the sub-region is consistent with site-centered order;
  one Mn sublattice shows large displacement amplitude while the other shows much smaller displacements.
 \textbf{b}, Periodic lattice displacement map in another sub-region shows displacements on both sublattices. 
  Further, the sublattice displacement amplitudes are not equal which indicates intermediate order. 
  \textbf{c}, Mn displacements (dots) and sinusoidal envelope function (line) in the site-centered phase. 
  \textbf{d}, Mn displacements (dots) and sinusoidal envelope function (line) in the intermediate phase. 
  }
  \label{F:Fig3}
\end{figure}
\addtocounter{figure}{-1}
\begin{figure} [t!]
    \caption{(continued) 
  \textbf{e,f}, Linecuts across the imaginary part of local Fourier transforms, $\mathfrak{Im}\{\mathfrak{F}(\mathbf{q}) \}$, along the pseudocubic directions $\mathbf{a^{*}_{pc}}$ and $\mathbf{c^{*}_{pc}}$.
  The site-centered phase (green) maintains $\pm\mathbf{q}$ symmetry whereas the intermediate phase (magenta) shows clear differences between the $\pm\mathbf{q}$ peaks, an indication that the latter phase breaks inversion symmetry.
 \textbf{g}, Large field-of-view map of periodic Mn displacements shows a transition from site-centered order (left side) to intermediate order (right side).
 The scale bar corresponds to 2 nm.
}
\end{figure}


The displacement pattern shown in Fig.~\ref{F:Fig3}a matches site-centered order. 
The first Mn sublattice has a large displacement amplitude (8.2(9) pm) while the second has a comparatively negligible amplitude (1.3(6) pm), similar to the group theory prediction of site-centered order.
The manganese displacements as a function of position, $\mathbf{r}$, are sinusoidal (Fig.~\ref{F:Fig3}c) and may be described by $\mathbf{\Delta}_{\text{Mn}}(\mathbf{r}) = \mathbf{A} \sin(\mathbf{Q}^{CO}.\mathbf{r} + \phi)$   
where $\mathbf{\Delta}_{\text{Mn}}(\mathbf{r})$, $\mathbf{A}$, $\mathbf{Q}^{CO}$, and $\phi$ are the displacement, amplitude and wavevectors, and the phase, respectively.
The phase determines the centering of the modulation relative to the Mn sites, with site- and bond-centered order corresponding to $\phi = n\pi$ ($n$ is an integer) and $\phi = n\pi/2$ ($n$ is an odd integer), respectively.
A deviation of the phase from these two limits corresponds to a state intermediate between site and bond order.
Fitting the displacements in the region shown in Fig.~\ref{F:Fig3}a further supports that the modulation is predominantly site-centered with $\phi=$ 0.04(1) $\pi$.
From this measurement, we thus confirm the presence of the prototypical site-centered state. 

Remarkably, we discover that within the same sample another region has a distinct displacement pattern:
prominent Mn displacements occur in both Mn sublattices (Fig.~\ref{F:Fig3}b).
Importantly, the displacements have different amplitudes on each sublattice, unlike the predicted pattern for purely bond-centered order.
By comparing to the group theory analysis of displacement patterns, the observed structure is consistent with a state which is intermediate between pure site and bond order ($Pm$ structure).  
Figure~\ref{F:Fig3}d shows the aggregated Mn displacements and their sinusoidal envelope.
The intermediate structure is readily apparent since the maximum of the sinusoidal function is not centered on Mn sites nor exactly at the middle.
The phase is $\phi$ = 0.35(1) $\pi$ and the mean displacement amplitudes on the first and second sublattices are 2.9(9) pm and 5.8(7) pm, respectively.
Given that STEM is a projection imaging technique, we ruled out the possibility that the intermediate state merely reflects a projection of stacked site-centered states along the beam direction.
This is achieved by analyzing both the contrast variations in the image and the patterns of displacements (see Methods section 3 for more details). 
Cryo-STEM mapping of lattice displacements therefore provides direct evidence for intermediate charge order in manganites, an observation with implications for other oxides in which the exact character of charge ordering remains unresolved.

An intriguing consequence of the overlap of site- and bond-centered order is that additional crystal symmetries may be broken \cite{Efremov2004}.
In oxides or charge-transfer salts, intermediate order is predicted to break inversion symmetry due to the formation of uncompensated dipoles, motivating proposals for unconventional ferroelectricity emerging from electronic order \cite{Efremov2004,VanDenBrink2008,cheong2007multiferroics,lopes2008new,lunkenheimer2012multiferroicity,senn2012charge}.
To determine whether the intermediate phase is indeed non-centrosymmetric, we compute local complex-valued Fourier transforms of the site-centered and intermediate regions.
The imaginary part, $\mathfrak{Im}\{\mathfrak{F}(\mathbf{q}) \}$, is sensitive to the odd component of the atomic resolution image and hence to the breaking of inversion symmetry \cite{Hamidian2012}.  
For a non-centrosymmetric structure, we therefore expect the crystalline Bragg peaks at $\pm \mathbf{q}$ positions to be inequivalent.
Figures~\ref{F:Fig3}e and f show integrated line cuts through $\mathfrak{Im}\{\mathfrak{F}(\mathbf{q}) \}$ along the pseudocubic $\mathbf{a}^{*}_{pc}$ and $\mathbf{c}^{*}_{pc}$ directions, respectively. 
In the site-centered phase (green), the imaginary component of the Bragg peaks has the same amplitude at $\pm \mathbf{q}$ coordinates for both pseudo-cubic directions, which indicates that inversion symmetry is maintained.
In contrast, there is a significant difference in the intermediate case (magenta), confirming that inversion symmetry is broken in that region.
The group theory analysis also tells a consistent story with the intermediate phase having a non-centrosymmetric ($Pm$) structure.
While intermediate charge order can in principle be detected though the switching of bulk polarization, the finite conductivity and nanoscale spatial inhomogeneity of manganites have precluded such an approach \cite{VanDenBrink2008}.
The atomic-scale mapping of intermediate order and the detection of the resultant broken inversion symmetry represent a new approach to probe electronically driven symmetry breaking.

Another key insight from these local visualizations is that ground states with distinct symmetries may coexist within the same system.
Both Figs. 3A and B are taken from the same field of view image (Fig. 3g), in which the site-centered phase (left side) transforms into the intermediate phase (right side) over a few unit cells. 
This suggests that the energies of site and intermediate orders are comparable and likely linked to subtle spatial fluctuations inherent to strongly interacting oxide materials \cite{lai2010mesoscopic}.
Such nanoscale coexistence further highlights the long-standing difficulty in determining the correct charge order model from bulk-averaged measurements, a situation exacerbated by the presence of crystalline twins (see for example Supplementary Fig. 3) and other inhomogeneity.

\begin{figure}
  \includegraphics[width=.65\linewidth]{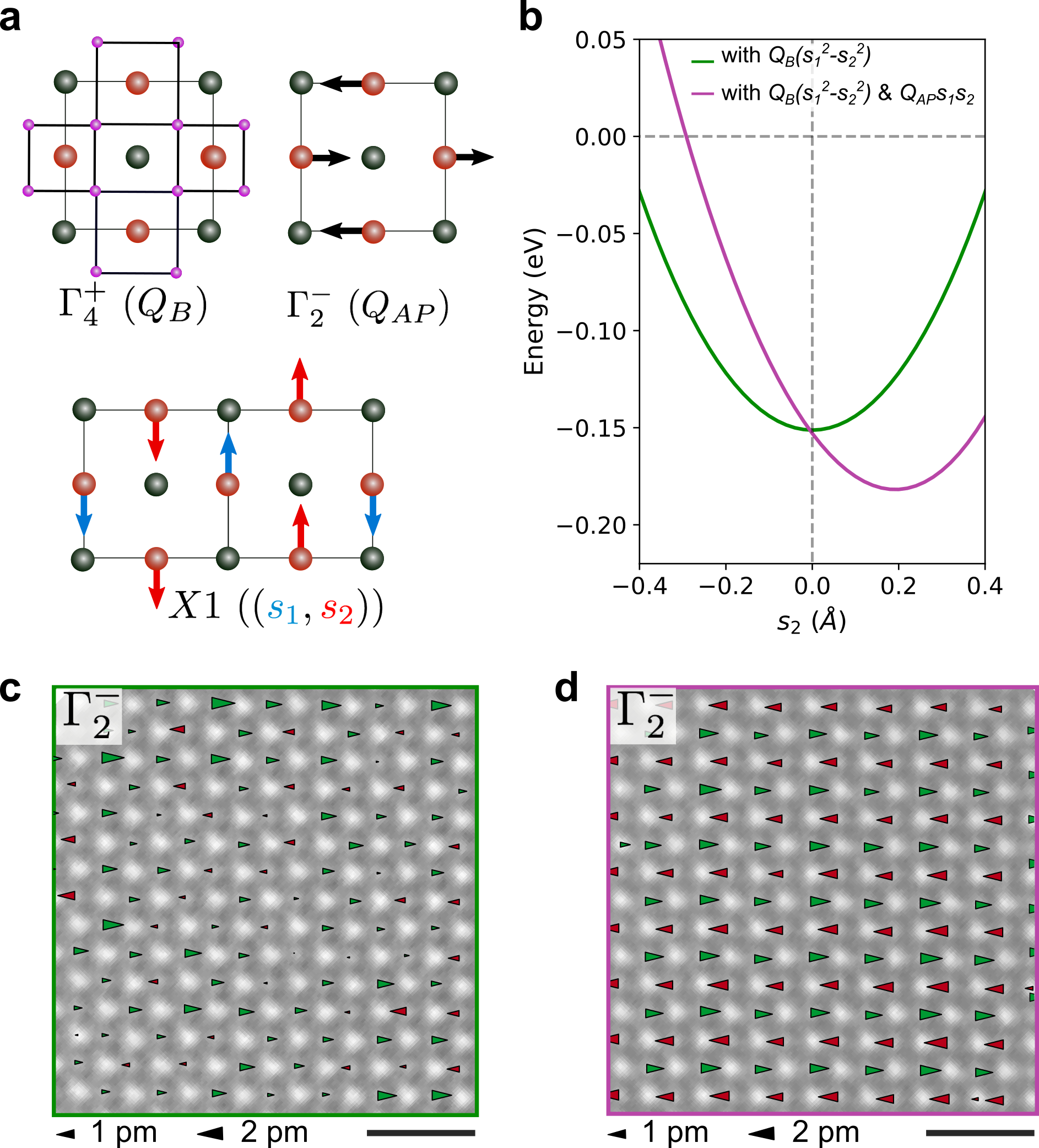}
  \caption{
  \textbf{Non-linear lattice coupling determines the charge order ground state.} 
  \textbf{a}, Displacement patterns for the $\Gamma_{4}^{+}$ mode (oxygen breathing distortion), the $\Gamma_{2}^{-}$ mode (Mn antipolar displacement), and the X1 mode.
  The purple dots represent the oxygen atoms.
  \textbf{b}, One dimensional plot of the full Landau free-energy (equation (3) in Methods section 5) along the ($0$, $s_{2}$) direction of X1.
  When the first coupling term is included and the second term is suppressed (green), the minimum energy is located at ($s_{1}$ $\neq 0$,$s_{2}$ = 0), indicating that site-centered order is favorable.
  If both coupling terms are present (magenta), both $s_{1}$ and $s_{2}$ are finite but not equal, indicating that intermediate order is favorable. 
  The values for the coefficients in the Landau free energy and the $Q_{AP}$/$Q_{B}$ amplitudes are listed in the Methods section 5.
  The amplitude of $s_2$ (in \AA) is with respect to a 40-atom DFT supercell.
  \textbf{c,d}, Atomic-scale maps of the $\Gamma_{2}^{-}$ Mn displacements in the site-centered and intermediate region, respectively.
  The largest arrows correspond to $\sim$ 2 pm.
  The displacements are disordered in the former and coherent in the latter, in agreement with the Landau theory predicting that the $\Gamma_{2}^{-}$ mode favors X1 displacements on both sublattices. 
  Both scale bars correspond to 1 nm.
  }
  \label{F:Fig4}
\end{figure}

Informed by the atomic-scale evidence for site and intermediate orders, we next examine the possible origin of these two states.
While we have focused on the X1 displacements so far, the low temperature phase may also contain additional structural responses that are allowed by symmetry (Fig.~\ref{F:Fig4}a). 
In particular, a $\Gamma_4^+$ mode, which appears as an oxygen breathing distortion, and a $\Gamma_2^-$ mode, which appears as antipolar displacements of Mn atoms, are such symmetry-allowed responses (see Methods section 4).   
To understand their role we expand the Landau free energy about the high-symmetry phase up to fourth order in powers of the $\Gamma_{4}^{+}$, $\Gamma_{2}^{-}$, and X1 displacement mode amplitudes (see equation (3) in Methods section 5 for the expansion).
Significantly, this expansion contains two relevant third-order coupling terms:
\begin{equation}
\label{e:couplmain}
F_{3} = \delta_{bss} Q_{B}(s_{1}^{2} - s_{2}^{2}) + \delta_{tss} Q_{AP}s_{1}s_{2}
\end{equation}
where $Q_{B}$, $Q_{AP}$, and ($s_{1}$, $s_{2}$) are the amplitudes of the $\Gamma_{4}^{+}$, $\Gamma_{2}^{-}$, and X1 distortions, respectively.
From density functional theory (DFT) calculations (see Methods section 5), we find that the $\delta_{tss}$ and $\delta_{bss}$ coefficients are negative, which indicates that the coupling terms lower the energy.
We therefore propose that the third-order nonlinear couplings between the $\Gamma$ and X1 modes provide a mechanism for favoring one ground state over another.

We now consider the implications of the two coupling terms in equation~(\ref{e:couplmain}). 
The first term stabilizes the breathing distortion and lowers the energy the most in the site-centered phase where the X1 displacements are ($s_1 \neq 0$, $s_2=0$). 
In contrast, the second term lowers the energy the most in the presence of the antipolar distortions and the bond-centered phase which has X1 displacements ($s_1$,$s_2=s_1$). 
Both coupling terms will contribute to lowering the energy in the intermediate case where the X1 amplitude is ($s_{1}$,$s_{2}\neq s_{1})$. 
To illustrate, we show one-dimensional line cuts across the minimum of the Landau energy in Fig.~\ref{F:Fig4}b, having set the coefficients and amplitudes to the DFT-calculated values (see Methods section 5 and Supplementary Tables II and III).
If we keep the first coupling term and suppress the second, the free-energy is minimized by having ($s_{1} \neq 0$,$s_{2}=0$) in the X1 amplitude, indicating that site-centered order is favorable.
When both coupling terms are present, the minimum of the free-energy shifts away from ($s_{1}$,$0$) to ($s_{1}$,$s_{2}\neq 0$), thus stabilizing the intermediate phase through a non-linear mechanism (see Supplementary Fig. 10 for alternative 2D plots of the energy surfaces).
Based on this theory, a key prediction is that the antipolar $\Gamma_{2}^{-}$ displacements are absent (present) in the experimentally observed site-centered (intermediate) phase.

To test this prediction, we visualize the spatial interplay between the antipolar Mn displacements and the character of the charge-ordered ground state.
Figures~\ref{F:Fig4}c and d show maps of the antipolar $\Gamma_{2}^{-}$ Mn displacements in the regions containing site and intermediate order, respectively. 
In the region with site-centered order, the $\Gamma_{2}^{-}$ mode is disordered, lacking any clear pattern of antipolar distortions.
In the region with intermediate order, however, this mode is coherent and relatively strong, in agreement with the Landau theory prediction.

The unusual non-linear third-order lattice couplings not only challenge our microscopic understanding of the origin of charge ordering but also provide a mechanism for manipulating the character of electronic order and its associated electronic properties. 
In the case of the non-centrosymmetric intermediate phase, an exciting prospect is to demonstrate ferroelectricity through the switching of polarization;
however, the finite conductivity of manganites and the manifestation of the intermediate phase only in localized regions are significant hurdles.  
To address the latter limitation, one approach inspired by our observations is to enhance the $Q_{AP}$ amplitude which in turn would favor the polar intermediate phase over macroscopic volumes. 
Such manipulation could be achieved via elastic strain and rare-earth substitution which were found to modulate the amplitude of $\Gamma_{2}^{-}$ distortions in related theoretical calculations \cite{nowadnick2019coupled}.  

Our study shows that cryogenic STEM is an emerging methodology for probing complex electronic ordering phenomena through the important but often neglected lens of the lattice degrees of freedom.
Using this approach, we discovered the co-existence of distinct charge order configurations including a pure site-centered phase and a more exotic intermediate phase which breaks inversion symmetry.
We also revealed that the stability of these different configurations depends on unique lattice couplings. 
Such a rich electronic and structural landscape should be relevant to other charge-ordered systems, including cuprates and nickelates where the exact microscopic arrangement associated with various electronic instabilities remains under intense scrutiny.
Similar to the current study, novel insights may be achieved through direct, real space visualizations that can spatially disentangle different ordering models and characterize the \textit{intra}-unit-cell structure and symmetry in detail.

\bibliography{refs.bib}
\bibliographystyle{custom}

\section*{Acknowledgments}
This work was primarily support by the Department of Defense Air Force Office of Scientific Research (FA 9550-16-1-0305). I.E.B. and L.F.K. acknowledge partial support by the National Sciences Foundation through the PARADIM Materials Innovation Platform (DMR-1539918).This work made use of the Cornell Center for Materials Research facilities supported through the NSF MRSEC program (DMR-1719875). The FEI Titan Themis 300 was acquired through NSF-MRI-1429155, with additional support from Cornell University, the Weill Institute and the Kavli Institute at Cornell.
E.A.N. acknowledges support from the University of California, Merced and the use of computational resources supported by the Cornell University Center for Advanced Computing.
The work at Stanford was supported by the the U.S. Department of Energy, Office of Basic Energy Sciences, Division of Materials Sciences and Engineering, 
under contract no. DE-AC02-76SF00515.
 \section*{Competing Interests}
The authors declare that they have no competing financial interests.
  \section*{Data availability} 
 The data that support the findings of this study are available from the corresponding author on reasonable request.
 \section*{Correspondence} 
 Correspondence and requests for materials should be addressed to L.F.K. \\ (email: lena.f.kourkoutis@cornell.edu) 
\section*{Author Contributions}
D.L., Y.H. and H.H. synthesized the samples and performed magnetic characterization.
D.J.B, I.E., M.J.Z. and L.F.K. acquired and analyzed the electron microscopy data.
E.A.N. performed density functional theory calculations.
E.A.N., I.E. and L.F.K. performed symmetry and theoretical analysis.
I.E., E.A.N., D.J.B. and L.F.K. wrote the manuscript with input from all authors.

\end{document}


\section*{Methods and Supplementary Materials}

\noindent
\textbf{1. Pulsed Laser Deposition Growth} To grow Nd$_{1/2}$Sr$_{1/2}$MnO$_{3}$ thin films, TiO$_{2}$-terminated SrTiO$_{3}$ (110) substrates were pre-annealed at 950$^{\circ}$C under oxygen partial pressure, p(O$_{2}$), of $5.0 \times 10^{-6}$ Torr for 30 min. 
The particular substrate orientation of SrTiO$_{3}$, namely the (110) orientation, is necessary for the charge order phase to emerge, since rotational anisotropy is required \cite{Nakamura2005}. 
Subsequently, 80-nm thick Nd$_{1/2}$Sr$_{1/2}$MnO$_{3}$ films were grown on top using a 248-nm KrF excimer laser. 
During the thin film deposition, the growth temperature was maintained at 825$^{\circ}$C under p(O$_{2}$) of $2.0 \times 10^{-3}$ Torr, using 0.40 J cm$^{-2}$ laser fluence and 11 mm$^{2}$ spot size.\\

\noindent
\textbf{2. Magnetization} The magnetization data were measured using a superconducting quantum interference device (SQUID) magnetometer with an in-plane magnetic field.
The sample was cooled down from room temperature to 10 K in zero external field, and its magnetization was measured in a 1000 Oe field while warming.
\\

\noindent
\textbf{3. Scanning transmission electron microscopy \& data analysis}\\
\noindent
\textbf{Experimental details.} Thin electron transparent samples were fabricated on a FEI Strata 400 focused ion beam using the standard liftout method with the stage tilted by 45$^{\circ}$. The stage tilt enabled preparation of TEM specimens in the correct crystal orientation, as the films are grown on (110)-oriented SrTiO$_{3}$ substrates.
High-angle annular dark-field (HAADF) STEM imaging was performed in an aberration-corrected FEI Titan Themis microscope operating at 300 kV and 50 pA probe current. 
The convergence semi-angle was 21.4 mrad and the inner collection angle was 68 mrad.
Cryogenic STEM measurements were performed using a Gatan double-tilt side-entry liquid nitrogen TEM holder (Model 636) with a base temperature of 93 K.
We acquired multiple fast-acquisition images (0.5 or 1 $\mu s$ per pixel and 1024$\times$1024 pixels per image) to overcome sample drift which occurs at low temperature, and aligned the image series using a rigid registration method optimized for noisy images~\cite{savitzky2018image}.

\begin{figure*}
  \centering
  \includegraphics[width=0.8\textwidth]{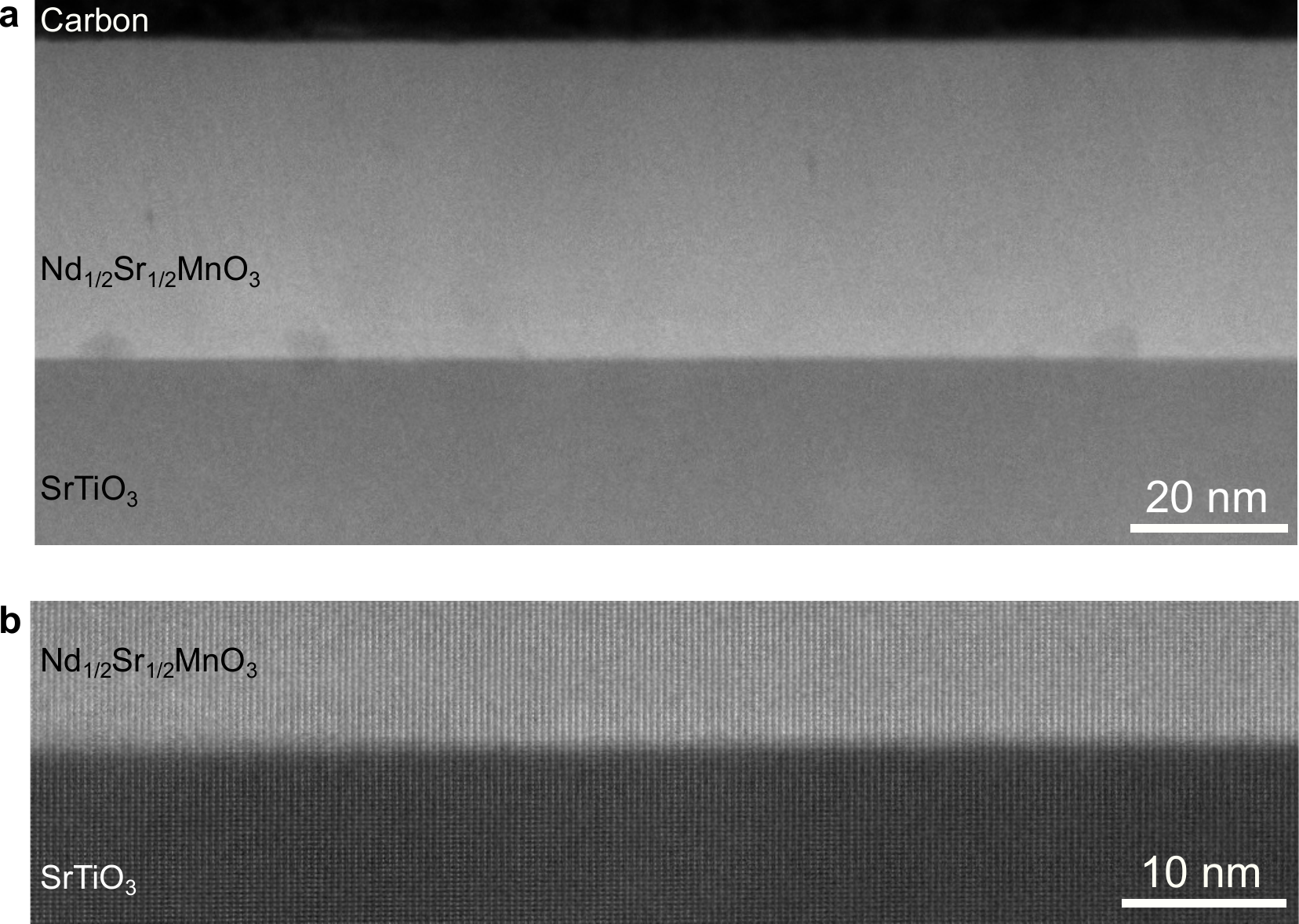}
  \caption{
      Thin film growth of Nd$_{1/2}$Sr$_{1/2}$MnO$_{3}$.
      \textbf{a}, Overview HAADF-STEM image of the sample. 
      A 80 nm thick film of Nd$_{1/2}$Sr$_{1/2}$MnO$_{3}$ is grown on a SrTiO$_{3}$ substrate using pulsed laser deposition.  
      \textbf{b}, Higher magnification STEM image showing the (110) orientation of the SrTiO$_{3}$ substrate and high-quality epitaxial growth of the Nd$_{1/2}$Sr$_{1/2}$MnO$_{3}$ film.
}
  \label{F:Interface}
\end{figure*}

\begin{figure*}
  \centering
  \includegraphics[width=0.8\textwidth]{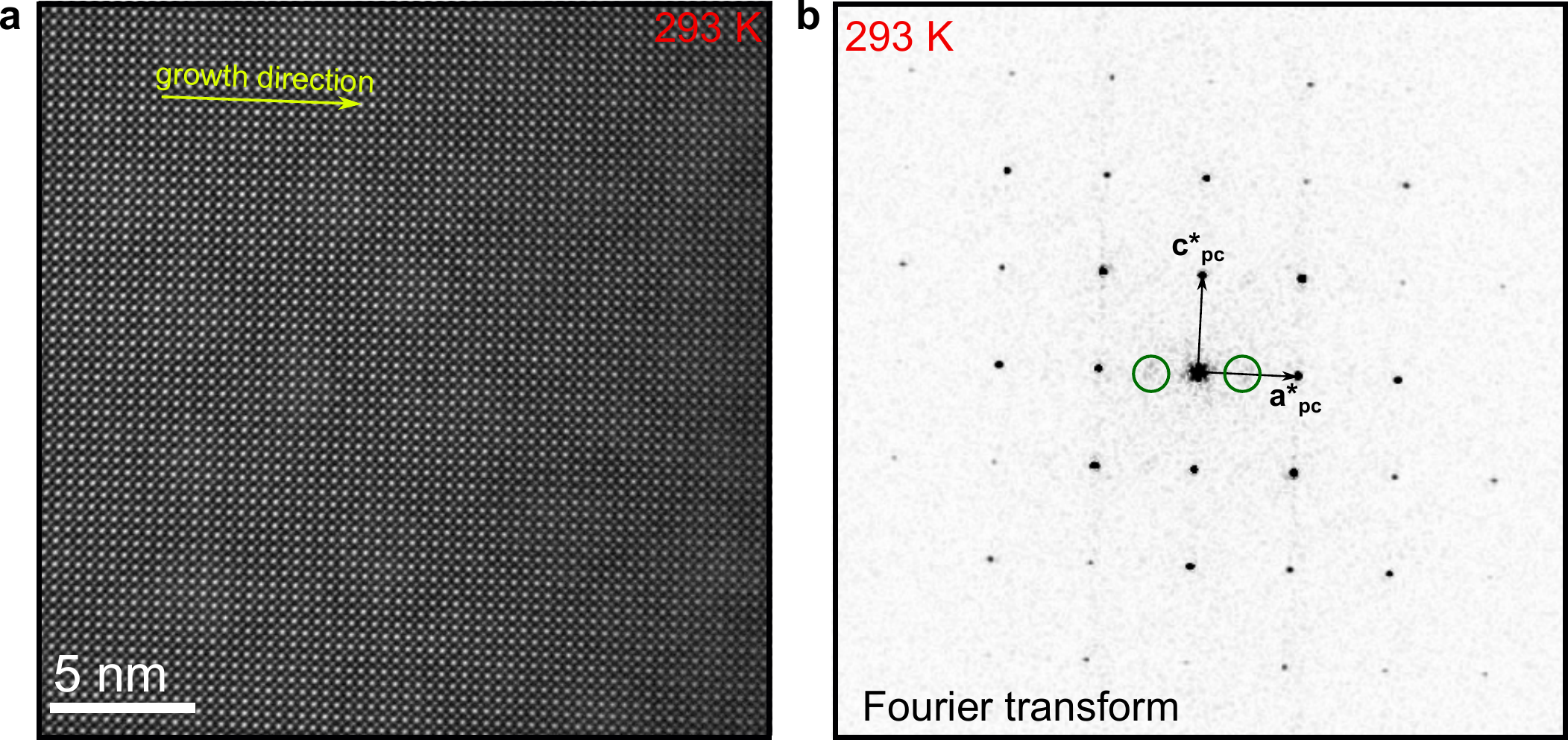}
  \caption{Room temperature data.
     \textbf{a}, HAADF STEM image of Nd$_{1/2}$Sr$_{1/2}$MnO$_{3}$ at room temperature.
     \textbf{b}, At this temperature, the Fourier transform amplitude shows no superlattice peaks associated with charge ordering, in agreement with the magnetization data (Fig. 1 in main text).
      Broad peaks associated with short-range cation order along the pseudocubic $\mathbf{a}^{*}_{pc}$-direction are present (green circles).
}
  \label{F:Room_Temperature}
\end{figure*}

\noindent
\textbf{Characterization of epitaxial film quality and fluctuations.}
\textit{Room temperature data.} The 80 nm epitaxial Nd$_{1/2}$Sr$_{1/2}$MnO$_{3}$/SrTiO$_{3}$ thin film is characterized using HAADF-STEM.
Supplementary Fig.~\ref{F:Interface}a shows a low magnification image that highlights the epitaxial growth of the film, with the substrate appearing dimmer than the film due to the atomic number contrast in HAADF-STEM.
At room temperature, no superlattice peaks associated with charge ordering are observed in the Fourier transform of STEM data (Supplementary Fig.~\ref{F:Room_Temperature}), in agreement with the magnetization data which shows that charge order occurs below 150 K. 
We note the presence of broad superlattice peaks (Supplementary Fig.\ref{F:Room_Temperature}b, green circles) which occur along the pseudocubic direction at $\mathbf{Q}^{\text{cat}}$ = ($\frac{1}{2},0,0$)$_{\text{pc}}$.
These peaks exist above and below the charge ordering transition temperature and are associated with local short-range A-site (Nd/Sr) cation ordering. 

\begin{figure*}
  \centering
  \includegraphics[width=0.9\textwidth]{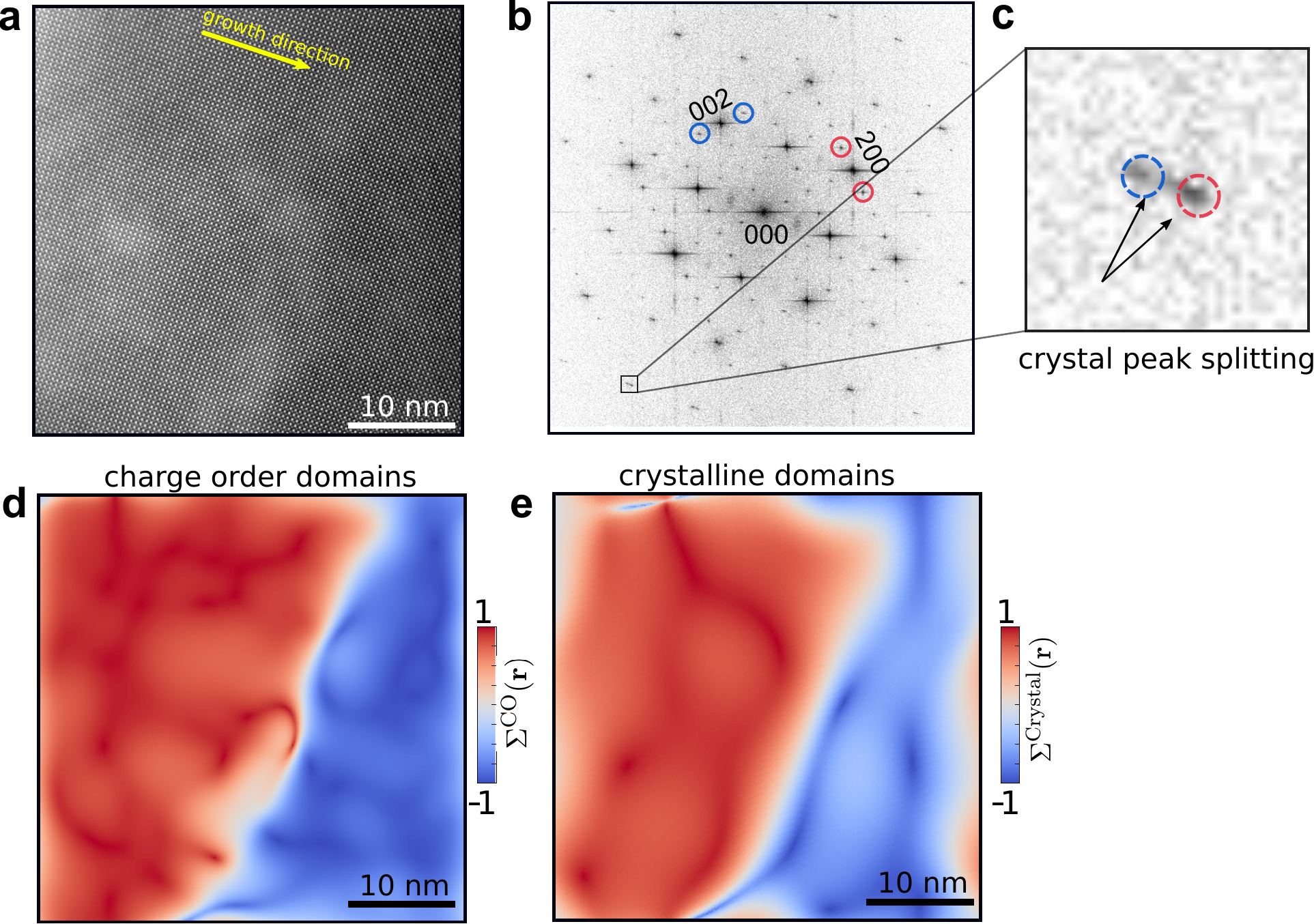}
  \caption{
      Crystal and charge-order domains are correlated.
      \textbf{a}, HAADF-STEM image over a large field of view (40 nm x 40 nm).
      \textbf{b}, Fourier transform amplitude highlighting orthogonal charge order peaks (blue/red circles).
      \textbf{c}, Zoom-in of the ($\bar{4}0\bar{2}$) crystal Bragg peak.
      A clear splitting is evident indicating the presence of crystalline orthorhombic twins.
     \textbf{d}, Local charge order unidirectionality parameter, $\Sigma^{\text{CO}}(\mathbf{r})$.
     A positive/negative value indicates that the contribution of the red/blue-highlighted charge order peak in \textbf{b} dominates locally. 
     \textbf{e}, Local crystal unidirectionality parameter, $\Sigma^{\text{Crystal}}(\mathbf{r})$,  mapped from the split ($\bar{4}0\bar{2}$) crystalline peaks (dashed circles).
      A positive/negative value indicates that the contribution of the red/blue-highlighted crystalline peak in \textbf{c} dominates locally. 
     The map shows a well-defined crystalline domain boundary.
     Further, this boundary tracks perfectly the charge domain boundary, indicating that the observed charge order twinning is due to crystal twinning and not to local nanoscale charge order clusters or overlapping domains.
  }
  \label{F:Crystal_Domain}
\end{figure*}

\textit{Crystal twins.} Structural inhomogeneities in bulk crystals, such as crystal twins and chemical fluctuations, often impede, or at the very least complicate, diffraction experiments since they require complex models or assumptions for structural refinements \cite{Woodward1999}.
Real space imaging, on the other hand, allows us to isolate and characterize local structural or chemical fluctuations in the manganite, from the occurrence of crystal and charge order domains to short-range cation ordering.
Supplementary Figure~\ref{F:Crystal_Domain}a shows a 40 nm region of the film in which we observe superlattice peaks along two orthogonal directions, $\mathbf{Q}^{CO}_{1}$ and $\mathbf{Q}^{CO}_{2}$, in the Fourier transform amplitude (Supplementary Fig.~\ref{F:Crystal_Domain}e).
In the case of bi-directional order, it is important to determine whether the modulations are overlapping or spatially separated \cite{comin2015broken,Savitzky2017}.
By Fourier filtering charge order superlattice peaks near the (200) and (002) Bragg reflections, we obtain the local amplitude of the modulation along each direction, $A_{\alpha}(\mathbf{r})$ with $\alpha$ = 1, 2.  
We define a local charge-order unidirectionality parameter
\begin{equation*}
    \Sigma^{\text{CO}}(\mathbf{r}) = [A_{1}(\mathbf{r}) - A_{2}(\mathbf{r})]/[A_{1}(\mathbf{r}) + A_{2}(\mathbf{r})]
\end{equation*}
which determines whether the ordering pattern in locally unidirectional (positive for the $\mathbf{Q}^{CO}_{2}$ direction, negative for the $\mathbf{Q}^{CO}_{1}$ direction and close to zero for overlapped).
The two charge order domains are separated by a well-defined boundary (Supplementary Fig.~\ref{F:Crystal_Domain}d), suggestive of a crystal twin effect rather than anomalous competition between unidirectional order parameters as seen in a previous study on a related system \cite{Savitzky2017}.
Indeed, we observe a splitting of crystalline Bragg peaks which indicates the presence of orthorhombic twins (Supplementary Fig.~\ref{F:Crystal_Domain}c).
A map of the local crystal unidirectionality parameter, $\Sigma^{\text{Crystal}}(\mathbf{r})$, determined using the split $(\bar{4}0\bar{2})$ Bragg peaks shows that it correlates with the charge order domains (Supplementary Fig.~\ref{F:Crystal_Domain}e).
The crystal twinning, and hence the charge order twinning, is likely due to the relaxation of the epitaxial film since it occurs along the growth direction.

\begin{figure*}
  \centering
  \includegraphics[width=0.99\textwidth]{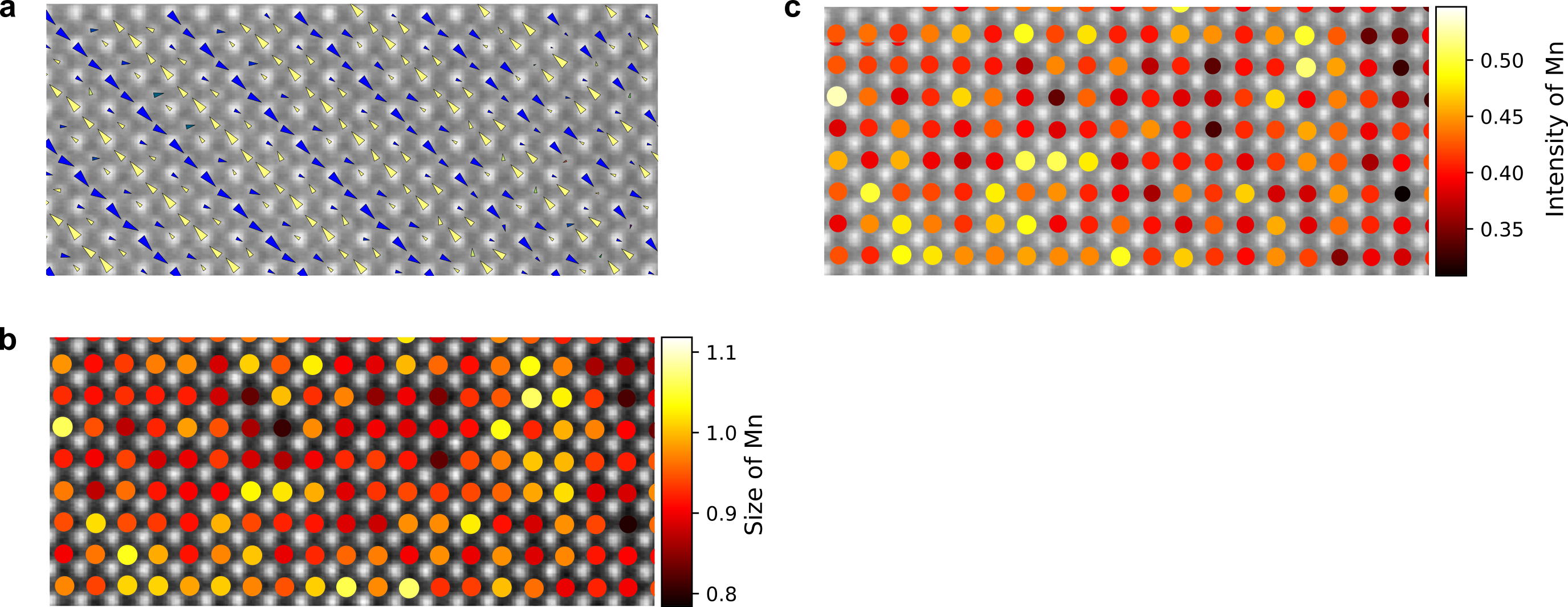}
  \caption{
      Ruling out stacking of site-centered order.
      \textbf{a}, Transition region containing regions with Mn displacements on one sublattice only (site-centered) and regions with Mn displacements on two sublattices. 
      If the latter emerges from out-of-plane stacking of site-centered domains, Mn columns would contain a combination of distorted and un-distorted atoms, leading to changes in HAADF-STEM intensity due to electron beam channeling.
      \textbf{b,c}, Gaussian fits to Mn columns in the transition region show no systematic variations of intensity or size of Mn columns.
      For instance, brighter/darker columns occur in both the site-centered regions and the regions with displacements on both Mn sublattices.
  }
  \label{F:Intensity}
\end{figure*}

\noindent
\textbf{Mapping primary displacements.}
The method for extracting periodic lattice displacements and rigorous numerical tests are described in great detail in previous work \cite{Savitzky2017,ElBaggari2018}.
To briefly summarize the method, we canvas the low temperature Fourier transform amplitude for all $\mathbf{Q}^{CO}=(1/2,0,0)$ superlattice peaks associated with charge ordering and damp their amplitude to the background level. 
We then apply an inverse Fourier transform, generating a reference image in which the modulation is removed.
The positions of the atomic columns are determined by fitting 2D Gaussian functions to STEM images \cite{Yankovich2014,Savitzky2017,ElBaggari2018}. 
Fitting and subtracting the lattice positions in the image pair, namely the original image and the reference image, resolves the structural modification which generates the superlattice peaks at low temperature.
The structural response corresponds to X1 displacements (see Methods section 4).  
\\
\noindent
\noindent
\textbf{Ruling out stacking of site-centered domains.} STEM imaging is a projection technique which averages out structural information along the beam direction.
We must therefore address the question of whether intermediate displacements are coherent throughout the volume or simply emerge from stacked, shifted domains of site-centered order. 
In the latter case, Mn columns will consist of a combination of distorted Mn atoms and non-distorted Mn atoms. 
Due to channeling effects in HAADF-STEM, the incoherent stacking of Mn atoms would result in distorted atomic column shapes and reduced intensity \cite{perovic1993imaging,hillyard1995detector,Fitting2006,Haruta2009}.
Supplementary Fig.~\ref{F:Intensity}a shows a transition region where both site and intermediate orders appear.
The small transition region ensures similar sample thickness.
Gaussian fits of the Mn atomic columns directly yield the intensity and size parameters on each column;
across the two charge order ground states, we observe no significant, systematic variation of either the column intensities or the column sizes (Supplementary Figs.~\ref{F:Intensity}b and c), indicating that in both cases Mn atoms are stacked coherently along the beam direction.

We can further confirm that the intermediate phase is a bona-fide ground state by comparing secondary structural responses within the site and intermediate regions.
The displacements in the intermediate phase cannot emerge from simple stacking of those in the site-centered phase.
For instance, we mapped the $\Gamma_{2}^{-}$ displacement pattern in the two regions (see Methods section 4). 
The site-centered phase shows suppressed and disordered $\Gamma_{2}^{-}$ displacements whereas the intermediate phase has coherent and prominent $\Gamma_{2}^{-}$ displacements (Supplementary Fig.~\ref{F:QTvX1}).
Incoherent stacking of the disordered displacements cannot a yield well-defined, periodic displacement pattern in projection. 
Therefore, the X1 displacements in the intermediate phase are intrinsic and coherent along the imaging direction and do not reflect a stacking of phase-shifted domains of site-centered displacements.\\
\noindent
\textbf{Mapping secondary displacements.}  The $\Gamma^{-}_{2}$ mode is a zone-center mode of the orthorhombic $Pnma$ cell and can be extracted using the same method if applied to the relevant peaks \cite{Savitzky2017,ElBaggari2018}.
To determine these peaks, we note that the $\Gamma^{-}_{2}$ modes (antipolar displacements of Mn atoms) correspond to zone-boundary modes of the cubic cell (Supplementary Fig.~\ref{F:QT}, dashed red square), therefore they have well defined and finite $\mathbf{q}$ peaks in the Fourier transform amplitude (Supplementary Fig.~\ref{F:QT}a, pink circles).
Indeed, these modes' periodicity is encoded in the $\mathbf{q}=(1,0,0)$-type Fourier peaks (Supplementary Fig.~\ref{F:QT}, pink circles).
To extract these displacements, we generate a reference image by removing the contribution of these particular peaks. 

The obtained displacements are shown in Supplementary Fig.~\ref{F:QT}c (full field of view) and e (zoom-in). 
The projection of the displacements along the $\Gamma_{2}^{-}$ distortion mode (see Methods section 4 for details about this mode and Supplementary Fig.~\ref{F:QT}d for the displacement pattern) is shown in Supplementary Fig.~\ref{F:QT}f.

Over the full field of view, the amplitude and coherence of this mode fluctuates in lock-step with the fluctuation in X1 displacements (Supplementary Fig.~\ref{F:QTvX1}).
In the site-centered region, $\Gamma_{2}^{-}$ displacements are disordered and weak (Supplementary Fig.~\ref{F:QTvX1}c and d).
However, they are strong and coherent in the intermediate region (Supplementary Fig.~\ref{F:QTvX1}e and f). 
As argued in the Landau theory (Methods section 4), the presence of the $\Gamma_{2}^{-}$ mode enhances the tendency to have X1 displacements on both Mn sublattices, stabilizing intermediate or bond-centered orders.\\
\noindent
\textbf{Fourier transforms.}
The complex-valued Fourier transforms, $\mathfrak{F}(\mathbf{q})$, in Figure 3 of the main text are taken from local regions with site order and intermediate order.
These regions are shown in Supplementary Fig.~\ref{F:QTvX1}.
Each region is multiplied by a 2D Hann window in order to removes contributions of sharp edges to the Fourier transforms.
We then take the imaginary part of the Fourier transform, $\mathfrak{Im}\{\mathfrak{F}(\mathbf{q}) \}$, which is sensitive to the odd component of the signal (image) and hence can be used to detect the breaking of inversion symmetry.
The real part, on the other hand, is sensitive to the even component of the signal and cannot determine inversion symmetry breaking.

A purely odd signal will have a purely imaginary and odd $\mathfrak{F(\mathbf{q})}$ while a purely even signal will have a purely real and even $\mathfrak{F(\mathbf{q})}$. 
An experimental image of a crystal structure, however, is much more complicated and has contributions from imperfect centering of the signal and phase offsets, which means that the Fourier transforms will have both real and imaginary components.
Since both charge-ordered structures occur within the same image, we can focus on the prominent differences in $\mathfrak{Im}\{\mathfrak{F}(\mathbf{q}) \}$ between the two models.
After taking the local Fourier transforms, we extract integrated line cuts across the Bragg peaks (pseudo-cubic directions).
The width of the integration window corresponds to the width of the Bragg peaks.\\

\noindent
\textbf{4. Landau Theory}\\
\noindent
\textbf{Subgroups.} Group theory analysis and symmetry decomposition were performed using the ISOTROPY software suite \cite{isotrophy,stokes2006isodisplace}.
First, we enumerate the possible distorted structures, or subgroups, that emerge from the high-symmetry structure in the $Pnma$ phase.
To double the unit cell along the [100] direction, we require a zone-boundary displacement mode with $\mathbf{q}$ = ($\frac{1}{2},0,0$).
There are only two such order parameters, with one transforming like the irreducible representation (irrep) $X1$ and the other like $X2$ (notation of Miller and Love \cite{miller1967tables,isotrophy}).
The subgroups that are generated by distinct directions of the X1 irrep are $P2_{1}/m$, $Pnm2_1$ and $Pm$ while those that can be generated by $X2$ are $P2_{1}/c$, $Pna2_{1}$ and $Pc$.\\
\noindent
\textbf{Displacement patterns.} We do not observe any $X2$-type displacements in the STEM data and no previous structural refinements have adequately described the charge-ordered structure in terms of the $X2$ displacement modes or their associated subgroups \cite{Radaelli1997,Daoud2002}.
Thus, the $X1$ mode is the relevant order parameter.
While this mode involves displacements of both cations and oxygen atoms, we will focus on the Mn atoms (sites).
For the Mn atoms, the $X1$ mode appears as periodic lattice displacements which can either be longitudinal, transverse and in-plane, or transverse and out-of-the plane.
Our data, as well as previous refinements, show that the transverse in-plane component dominates;
therefore we represent the X1 mode in terms of the transverse in-plane component, as done in Figs. 1 and 2 in the main text. 

As discussed in the main text, the $X1$ order parameter is two-dimensional with each dimension corresponding to a Mn sublattice in the $Pnma$ cell.
The $P2_{1}/m$ subgroup is generated by an X1 distortion along a single dimension ($s_1$,$s_2$=0); 
$Pnm2_1$ by X1 distortions along both dimensions and with equal magnitude ($s_1$,$s_2$=$s_1$);
and $Pm$ by X1 distortions along both dimensions but with different magnitude ($s_1$,$s_2 \neq s_1$).\\
\noindent
\textbf{Energy expansion and secondary modes.} We performed the free energy expansion about the high-symmetry phase (space group $Pnma$) in terms of polynomials of the X1 order parameter along the general direction, ($s_1$,$s_2$), up to fourth order \cite{hatch2003invariants}.
Importantly, we also include the secondary modes that are allowed by symmetry \cite{hatch2001complete,stokes2002copl}.
The possible secondary order parameters are limited to $\Gamma$ modes with $\mathbf{q}=(0,0,0)$ with respect to the $Pnma$ cell, and include $\Gamma_{1}^{+}$, $\Gamma_{4}^{+}$, $\Gamma_{2}^{-}$, and $\Gamma_{3}^{-}$.
The $\Gamma_{1}^{+}$ mode is a strain or volume change which appears, by symmetry, in any phase transition.
The $\Gamma_{4}^{+}$ mode is a displacement mode appearing as breathing distortion of the oxygen octahedra, which, from a chemical bond perspective, corresponds to charge localization \cite{brown1992chemical,nowadnick2019coupled}.
Note that this "charge localization" mode alone does not double the unit cell so it must require coupling to the X1 mode.
Further, according to DFT calculations, this mode by itself raises the energy of the system and only acquires a finite but small amplitude by coupling to other modes (see Methods section 5 and ref. \cite{nowadnick2019coupled}).
The $\Gamma_{2}^{-}$ mode involves predominantly oxygen displacements as well as antipolar displacements of the metal cations.
Since only the cations are visible in HAADF-STEM we will focus on the Mn displacement pattern associated with the $\Gamma_{2}^{-}$ irrep.
Finally, $\Gamma_{3}^{-}$ corresponds to a polar displacement mode.

The free energy will contain the usual quadratic and quartic terms for each of these listed order parameters (X1 and all the listed $\Gamma$ modes) and, more importantly, coupling terms whose form depends on the symmetry of the order parameters.
The full coupling energy is given by
\begin{equation}
\label{e:coupl00}
\begin{aligned}
F_{3} = & \delta_{vss} Q_{V}(s_{1}^{2} + s_{2}^{2}) + \delta_{bss} Q_{B}(s_{1}^{2} - s_{2}^{2}) + \\ 
& \delta_{tss} Q_{AP}s_{1}s_{2} + \delta_{btp}Q_{B}Q_{AP}Q_{P} 
\end{aligned}
\end{equation}
where $Q_{V}$, $Q_{B}$, $Q_{AP}$, Q$_{P}$, and ($s_{1}$, $s_{2}$) are the amplitude of the $\Gamma_{1}{+}$, $\Gamma_{4}^{+}$, $\Gamma_{2}^{-}$, $\Gamma_{3}^{-}$, and X1 distortions, respectively; 
and $\delta_{vss}$,  $\delta_{bss}$, $\delta_{tss}$ $\delta_{btp}$ are the coupling coefficients.
The coefficients can be determined using DFT calculations (see Methods section 5 and Supplementary Table~\ref{tab:coeff}).
We find that $\delta_{bss}$, $\delta_{tss}$ and $\delta_{btp}$ are negative, which means that the corresponding coupling terms lower the energy. 
\\
\noindent
\textbf{Interpretation of the coupling terms.} The first coupling term, $Q_{V}(s_{1}^{2} + s_{2}^{2})$, reflects the coupling of strain and volume change to the phase transition.
Moreover, the $\Gamma_{1}^{+}$ mode maintains the symmetry of the parent $Pnma$ structure and does not favor a particular direction in the X1 order parameter. 
Its main effect is quantitative, in that it slightly alters the coefficients of the usual $s_1$/$s_2$ quadratic terms in the Landau energy expansion. 

The interplay between the second and third coupling terms, on the other hand, is important for determining the nature of the ground state, as discussed in the main text.
The second term induces the $\Gamma_{4}^{+}$ ($Q_{B} \neq 0$) distortion and a ($s_{1}$, $0$) amplitude in the X1 mode, which yields site-centered order with $P2_{1}/m$ symmetry.
The third term induces the $\Gamma_{2}^{-}$ ($Q_{AP} \neq 0$) distortion and a ($s_{1}$, $s_{1}$) X1 amplitude, which yields bond-centered order with $Pnm2_1$ symmetry.
Only when both terms occur do we obtain the intermediate phase with ($s_{1}$, $s_{2}\neq s_{1}$) and $Pm$ symmetry.
In this case, both $\Gamma_{4}^{+}$ and $\Gamma_{2}^{-}$ are present ($Q_{B} \neq 0$ and $Q_{AP} \neq 0$).
Plots of the energy surfaces in the cases where (a) no coupling terms are included, (b) only the second term is non-zero, (c) only the third term is non-zero and (d) both terms are non-zero are shown in Supplementary Fig.~\ref{F:Landau}.

A consequence of the intermediate phase, where $Q_{B}$, $Q_{AP} \neq 0$,  is that the fourth term, $Q_{B}Q_{AP}Q_{P}$, can become finite if a finite polarization ($Q_{P} \neq 0$) appears, breaking inversion symmetry and lowering the energy. 
That the polar state appears due to the trilinear $Q_{B}Q_{AP}Q_{P}$ coupling term suggests that ferroelectricity, if switchable, should be categorized as hybrid improper ferroelectricity \cite{benedek2011hybrid,pitcher2015tilt}.
This analysis thus provides an explicit mechanism for the emergence of the polar state in the intermediate phase and enhances the proposal by Efremov \textit{et al.} \cite{Efremov2004}.

Finally, we note that in the main text we mainly focused on the second and third terms in $F_{3}$ because they are directly related to determining the stability of site and intermediate orders.
That said, the other coupling terms should be considered if a more quantitative analysis is required, especially in future studies of electric field or strain effects on charge order.\\

\noindent
\textbf{5. Density Functional Theory}\\
\noindent
\textbf{Computational details.}
We perform density functional theory (DFT) + $U$ calculations as implemented in the Vienna \textit{ab~initio} simulation package (VASP) \cite{kresse1993ab,Kresse1999} with the PBEsol exchange-correlation functional \cite{perdew2008restoring} and a 500 eV energy cutoff for the plane-wave basis. 
We perform our calculations in an orthorhombic 40 atom unit cell with dimensions $2\sqrt{2} \times 2 \times \sqrt{2} $  and choose a $4\times 6\times 8$ $\Gamma$-centered $k$-point grid.  
We use the Liechtenstein formulation of DFT + $U$ \cite{liechtenstein1995density}, where we set $U = 5$ eV and $J = 1.2$ eV. 
For structural relaxations, we use a 2.5 meV/\AA~ force convergence tolerance.

The starting point of the calculation is the high-symmetry structure $P4/mmm$ (which would correspond to the cubic cell without cation order) with Nd/Sr cations ordered along the unique \textbf{b}-axis. 
We perform calculations with ferromagnetic (FM), A-type antiferromagnetic (A-AFM), and CE-type antiferromagnetic (CE-AFM) orders imposed.
Note that for the CE-AFM case, we use an 80 atom cell with a $4\times6\times4$ $k$-point grid. 
The AFM-coupling in the A- and CE-AFM states is imposed along the \textbf{b}-axis.
We find that the polar $P2_1ma$ structure (intermediate phase) has the lowest energy (Supplementary Table~\ref{tab:structures}).
Note that in the DFT-calculated structures, the computational requirement for A-site (Nd/Sr) ordering changes the symmetry.

\noindent
\textbf{Displacement modes.}
To assess the symmetries and amplitudes of the structural distortions that contribute to the $P2_1ma$ structure, we report in Supplementary  Table~\ref{tab:decomp001} a structural decomposition of the DFT-relaxed $P2_1ma$ structure into symmetry adapted modes of the $P4/mmm$ reference structure \cite{isotrophy,stokes2006isodisplace}.
Note that since we are using a $P4/mmm$ (rather than $Pnma$) reference structure here, the labelling of the irreps is different, although they represent the same physical atomic displacements. The M$_{5}^{-}$ mode, which is an octahedral tilting mode (contained in $Pnma$)  together with antipolar Mn displacements, and the $\Sigma_{2}$ mode, which corresponds to the X1-type periodic lattice displacements, have the largest amplitudes, followed by the M$_{4}^{+}$  oxygen breathing distortion.
The polar ($\Gamma_{5}^{-}$) distortion is also present.
For each distortion, we use the displacement of each distinct atomic type to construct the relevant symmetry-adapted mode vector, which we use for our energy surface calculations below \cite{nowadnick2019coupled}.

\noindent
\textbf{Landau expansion about the $P4/mmm$ reference structure.}
With the $P4/mmm$ irreps defined above, we perform the energy expansion of the free energy, $F_1$, about the high-symmetry $P4/mmm$ reference structure \cite{stokes2002copl}.   
To fourth order, the energy is 
\begin{eqnarray}
\label{e:fulllandau}
F_1 &=& \tfrac{1}{2}{\alpha_t}Q_{AP}^2 + \tfrac{1}{4}{\beta_t} Q_{AP}^4 \\ 
&+&\tfrac{1}{2}{\alpha_s}(s_1^2+s_2^2)+ \tfrac{1}{4}{\beta_s}(s_1^2+s_2^2)^2  + \tfrac{1}{2}{\gamma_s}s_1^2s_2^2  \nonumber \\
& + & \tfrac{1}{2}{\alpha_b}Q_{B}^2 + \tfrac{1}{2}{\alpha_p}Q_{P}^2 \nonumber \\
& + & F_{3}\nonumber
\end{eqnarray}
where $F_3$ is the third-order coupling energy
\begin{equation}
\label{e:coupl}
   F_3 = \delta_{tss}Q_{AP}s_1s_2 + \delta_{bss}Q_{B}(s_1^2-s_2^2) +\delta_{btp}Q_{AP} Q_{B}Q_{P}.
\end{equation}
Here, $Q_{AP}$, $Q_{B}$, $Q_{P}$, and ($s_1$,$s_2$) are the amplitudes of the M$_{5}^{-}$, M$_{4}^{+}$, $\Gamma_{5}^{-}$ and $\Sigma_{2}$ displacement modes, respectively (see Table~\ref{tab:decomp001} in Methods section for the amplitudes). 
Note that the coupling terms in $F_3$ have the same functional form as in equation~(\ref{e:coupl00}) in Methods section 4.
\\
\noindent
\textbf{Determining the coupling coefficients.}
We calculate energy surfaces by ``freezing in" increasing amplitudes of each mode and calculating the total energy with DFT, as shown in Supplementary Fig.~\ref{fig:surf1}. 
By fitting these energy surfaces we extract the Landau expansion coefficients $\alpha_i$ and $\beta_i$ in equation~(\ref{e:fulllandau}). 
Note that we freeze in the two-dimensional $\Sigma_2$ order parameter along the direction (0.907, 0.421), which is the direction this order parameter takes in the DFT-relaxed $P2_1ma$ structure with A-AFM order. 
That is, we freeze in amplitude $Q_s=\sqrt{s_1^2+s_2^2}$ where $s_1 = 0.907Q_s$ and $s_2 = 0.421Q_s$. For this reason, we omit the $\gamma_s$ coefficient from equation~(\ref{e:fulllandau}).

To calculate the coupling coefficients $\delta_{tss}$ and $\delta_{bss}$ in equation~(\ref{e:coupl}), we fix the $\Sigma_2$ distortion amplitude to the value obtained in the DFT-relaxed structure ($Q_s = 0.68$ \AA), freeze in the $M_5^-$ distortion ($M_4^+$ breathing distortion), and then fit the resulting energy surface  (Supplementary Fig.~\ref{fig:surf2}a,b). 
To calculate the trilinear $\delta_{tbp}$ coefficient, we fix the $M_5^-$ and $M_4^+$ amplitudes to the values in the DFT-relaxed structure and freeze in the polar mode (Supplementary Fig.~\ref{fig:surf2}c).
The resulting coupling coefficients are reported in Supplementary Table~\ref{tab:coeff}.

The $\alpha_{s}$, $\alpha_{b}$, $\alpha_{p}$ coefficients are all positive which means that the X1-like displacements, the breathing distortion, and the polar distortion by themselves do not lower the energy.
The breathing distortion and the polar distortion significantly raise the energy so we limited the energy expansion to the second powers of $Q_{B}$ and $Q_{P}$.
Meanwhile, the coupling coefficients $\delta_{tss}$, $\delta_{bss}$ and $\delta_{btp}$ are negative, which means that the coupling terms lower the energy of the system by non-linearly inducing distortions that by themselves are energetically unfavorable.
The energy surfaces and line plots in Supplementary Fig.~\ref{F:Landau} and Fig. 4 of the main text are generated from supplementary equations~(\ref{e:fulllandau})-(\ref{e:coupl}) and Supplementary Table~\ref{tab:coeff}.
These results show that non-linear lattice coupling are essential for understanding the formation of the charge-ordered states.


\bibliography{methods.bib}
\bibliographystyle{custom}

\pagebreak

\begin{table}[]
\begin{center}
\caption{\label{tab:structures} Selected structural phases of 010 cation-ordered NdSrMn$_2$O$_6$. For each subgroup and magnetic order, structural relaxations were performed with DFT + $U$. Energies are reported for a 40 atom cell. The first two columns indicate the direction of the $M_5^-$ and $\Sigma_2$ irreps in each subgroup.}
\begin{tabular*}{\linewidth}{c @{\extracolsep{\fill}} ccccc}
\hline
\hline
$M_5^-$ & $\Sigma_2$ & Space group & Magnetic state & Energy [meV/f.u.] \\ 
\hline
- & - & $P4/mmm$ & A-AFM &0.0  \\ 
- & - & $P4/mmm$ & FM &-19.5  \\ 
\hline
($a$,0) & - & $Pmam$ & A-AFM & -121.3  \\ 
($a$,0) & - & $Pmam$ & FM & -126.6  \\ 
($a$,$a$) & - & $Cmmm$ & A-AFM  &-133.6  \\ 
($a$,$a$) & - &$Cmmm$ & FM &-138.2  \\ 
\hline
- & ($a$,$b$,0,0) & $P2_1ma$ & CE-AFM  &-161.3  \\ 
- & ($a$,$b$,0,0) &$P2_1ma$ & A-AFM & -154.7  \\ 
\hline
\end{tabular*}
\end{center}
\end{table}

\clearpage

\begin{table}[]
\caption{\label{tab:decomp001} Decomposition of the DFT+$U$-relaxed intermediate phase ($P2_1ma$) structure into symmetry adapted modes of $P4/mmm$. 
Here 010 cation order and A-type AFM order is imposed in the calculations. Amplitudes are given in units of \AA~for the 40 atom unit cell. The relaxed lattice vectors for the $P2_1ma$ cell are $a$=11.001 \AA, $b$ = 7.468 \AA, and $c$ = 5.445 \AA. The description shows the corresponding irrep in the case of no cation order (Methods section 4). 
\\}
\begin{tabular*}{\linewidth}{c @{\extracolsep{\fill}} cccc}
\hline
\hline
Irrep & Wavevector & Amplitude (\AA) & Description\\
\hline
$M_5^-$ & ($\frac{1}{2}$, $\frac{1}{2}$, 0) & 1.23 &  Antipolar Mn displacements ($\Gamma_2^-$)  \\
$M_4^+$ &  ($\frac{1}{2}$, $\frac{1}{2}$, 0) & 0.17 & Oxygen breathing distortion ($\Gamma_4^+$) \\
$\Sigma_2$ & ($\frac{1}{4}$, $\frac{1}{4}$,0) & (0.907, 0.421) & X1-like periodic displacements              \\
$\Gamma_5^-$ & (0,0,0) & 0.05 &                         Polar distortion ($\Gamma_5^-$)\\
\hline
\end{tabular*}
\end{table}

\clearpage

\begin{table}[]
\caption{\label{tab:coeff} Parameters from fitting the free-energy expansion in equations~(\ref{e:fulllandau})-(\ref{e:coupl}) to the DFT-calculated energy surfaces shown in Supplementary Fig.~\ref{fig:surf1}.\\}
\begin{tabular*}{\linewidth}{l @{\extracolsep{\fill}} lll}
\hline
\hline
$\alpha_{t} = -0.31$  eV/\AA$^{2}$   &  $\beta_{t}= 0.28$ eV/\AA$^{4}$    & $\delta_{tss}= -0.31$ eV/\AA$^{3}$   \\
$\alpha_{s} =  0.17$  eV/\AA$^{2}$   &  $\beta_{s}= 1.01$ eV/\AA$^{4}$    &  $\delta_{bss}= -2.14$ eV/\AA$^{3}$  \\
$\alpha_{b} =  2.72$  eV/\AA$^{2}$   &                                    &  $\delta_{btp}= -0.24$ eV/\AA$^{3}$  \\
$\alpha_{p} =  0.88$  eV/\AA$^{2}$   &                                    &                                      \\
\hline
\end{tabular*}
\end{table}

\clearpage

\clearpage

\clearpage

\clearpage
\begin{figure*}
  \centering
  \includegraphics[width=0.999\textwidth]{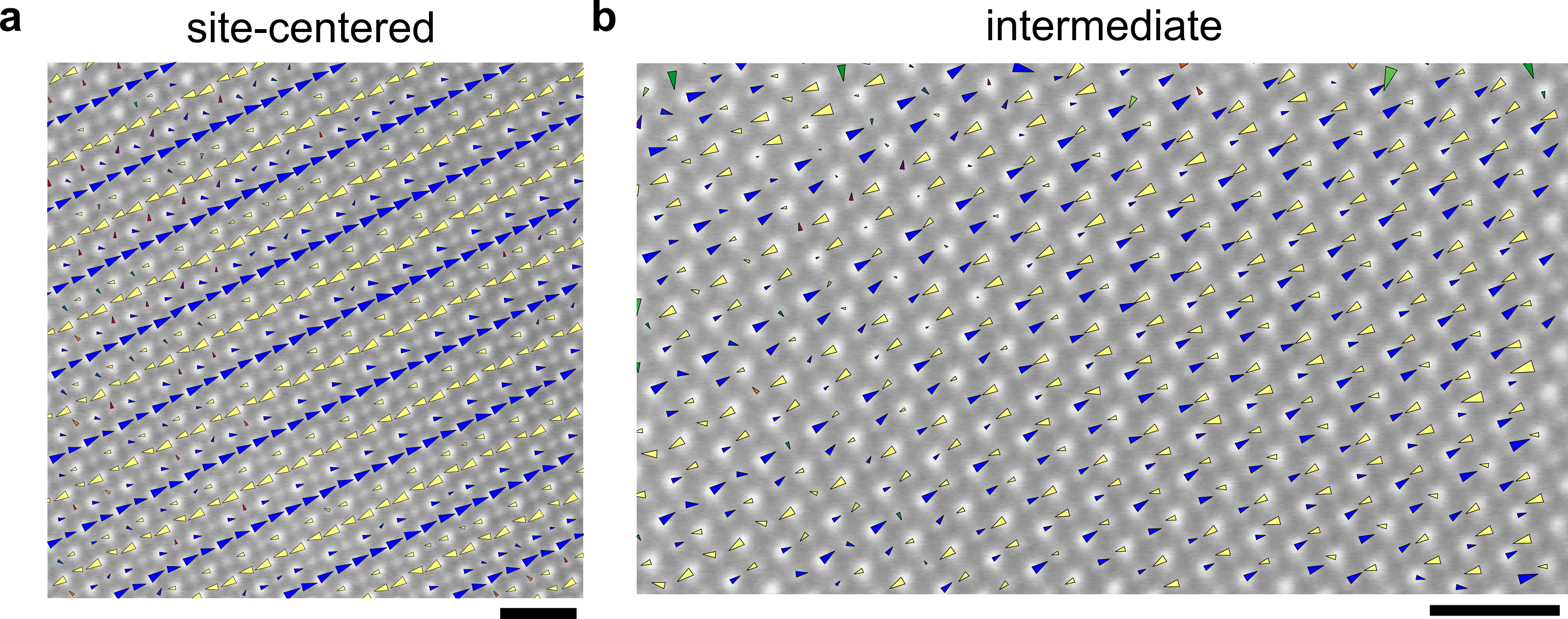}
  \caption{
        Additional data showing regions of site or intermediate order.
\textbf{a,b}, Site-centered order and intermediate order in regions distinct from those in the text.
Both the Mn and Nd/Sr displacements are shown here.
The scale bars correspond to 1 nm.  
}
  \label{F:AdditionalFOV}
\end{figure*}

\begin{figure*}
  \centering
  \includegraphics[width=0.8\textwidth]{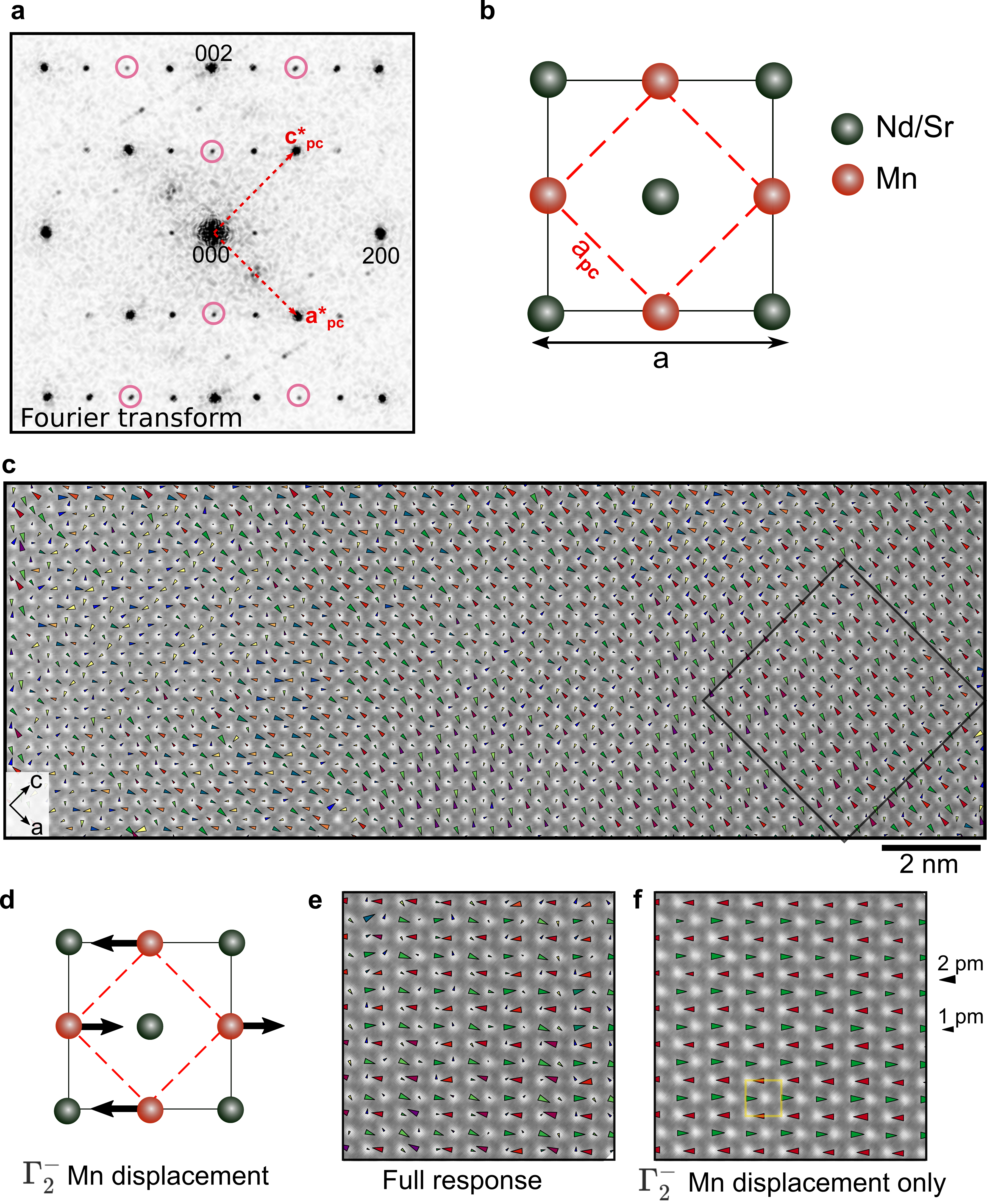}
  \caption{
      Extracting $\Gamma_{2}^{-}$ displacements.
      \textbf{a}, Fourier transform amplitude in the low temperature phase.
      Peaks associated with the $\Gamma$ modes of the orthorhombic cell are highlighted by pink circles. 
      The indexing of the peaks assumes a $Pnma$ structure. 
      \textbf{b}, Relationship between the orthorhombic cell (black square) and the pseudocubic cell (red dashed square).  
      \textbf{c}, Map of the lattice displacements associated with the orthorhombic peaks (pink circles in a).
      The Nd/Sr displacements are negligible.
      \textbf{d}, Displacement pattern of the $\Gamma_{2}^{-}$ mode on Mn sites.
      These displacements fit in the orthorhombic cell.
      \textbf{e}, Zoom-in of the region within the black square in \textbf{c}. 
      \textbf{f}, Projection of the extracted Mn displacements onto the $\Gamma_{2}^{-}$ mode direction.
  }
  \label{F:QT}
\end{figure*}

\begin{figure*}
  \centering
  \includegraphics[width=0.7\textwidth]{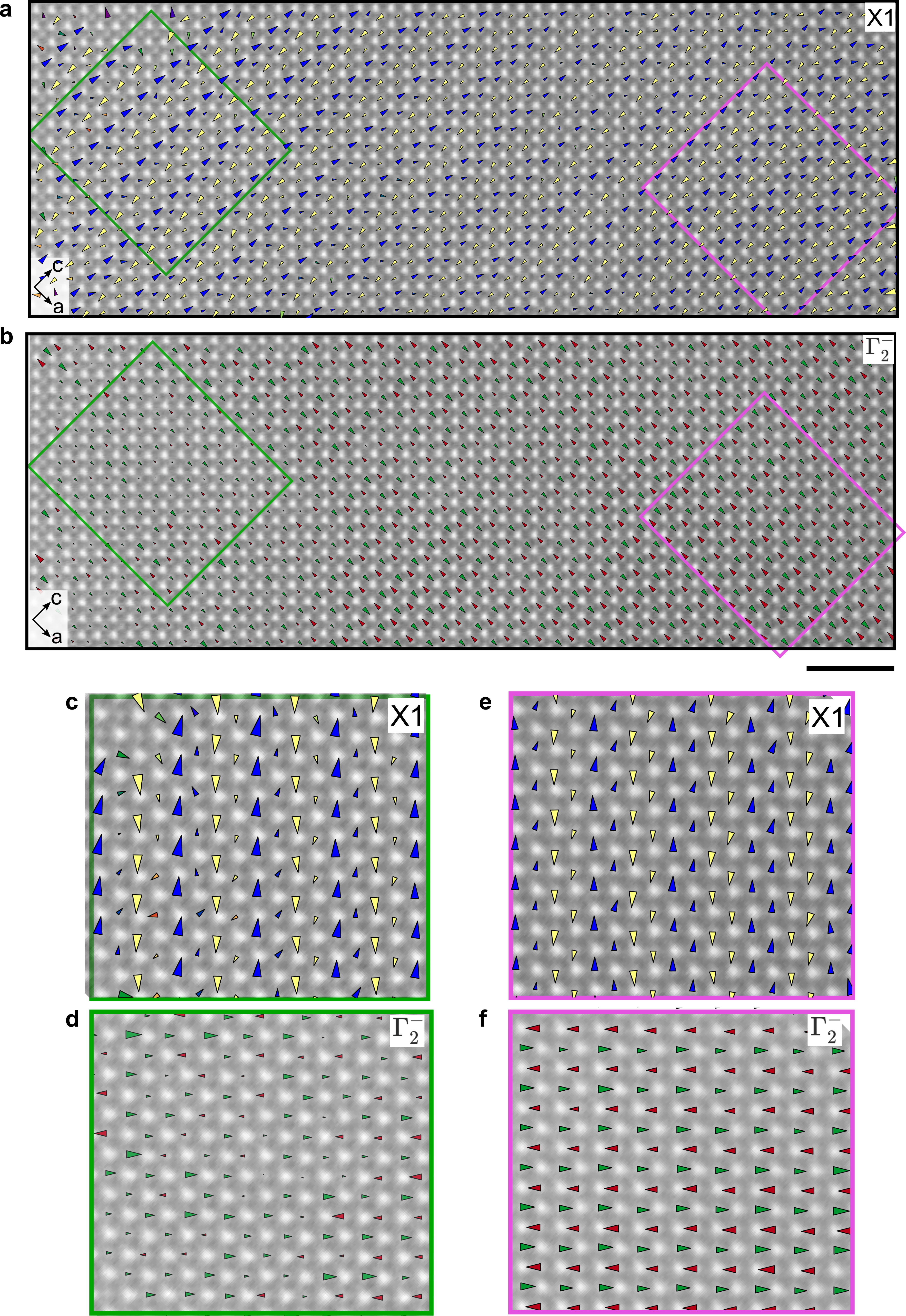}
  \caption{
        Interplay between $\Gamma_{2}^{-}$ displacements and $X1$ displacements.
\textbf{a}, Large field-of-view of X1 Mn displacements that transition from a site-centered (green) state to an intermediate (purple) state. 
The regions marked by the rectangles span those shown in Fig. 3 in the main text.
\textbf{b}, Map of $\Gamma_{2}^{-}$ Mn displacements in the same region as a.
\textbf{c,d}, Map of the X1 and $\Gamma_{2}^{-}$ Mn displacements within the site-centered region.
\textbf{e,f}, Map of the X1 and $\Gamma_{2}^{-}$ Mn displacements within the intermediate region.
The $\Gamma_{2}^{-}$ displacements are weak and disordered within the region containing site-centered order but are strong and coherent within the region containing intermediate order, in agreement with the Landau theory prediction. 
}
  \label{F:QTvX1}
\end{figure*}

\begin{figure*}
\begin{center}
\includegraphics[width=0.75\textwidth]{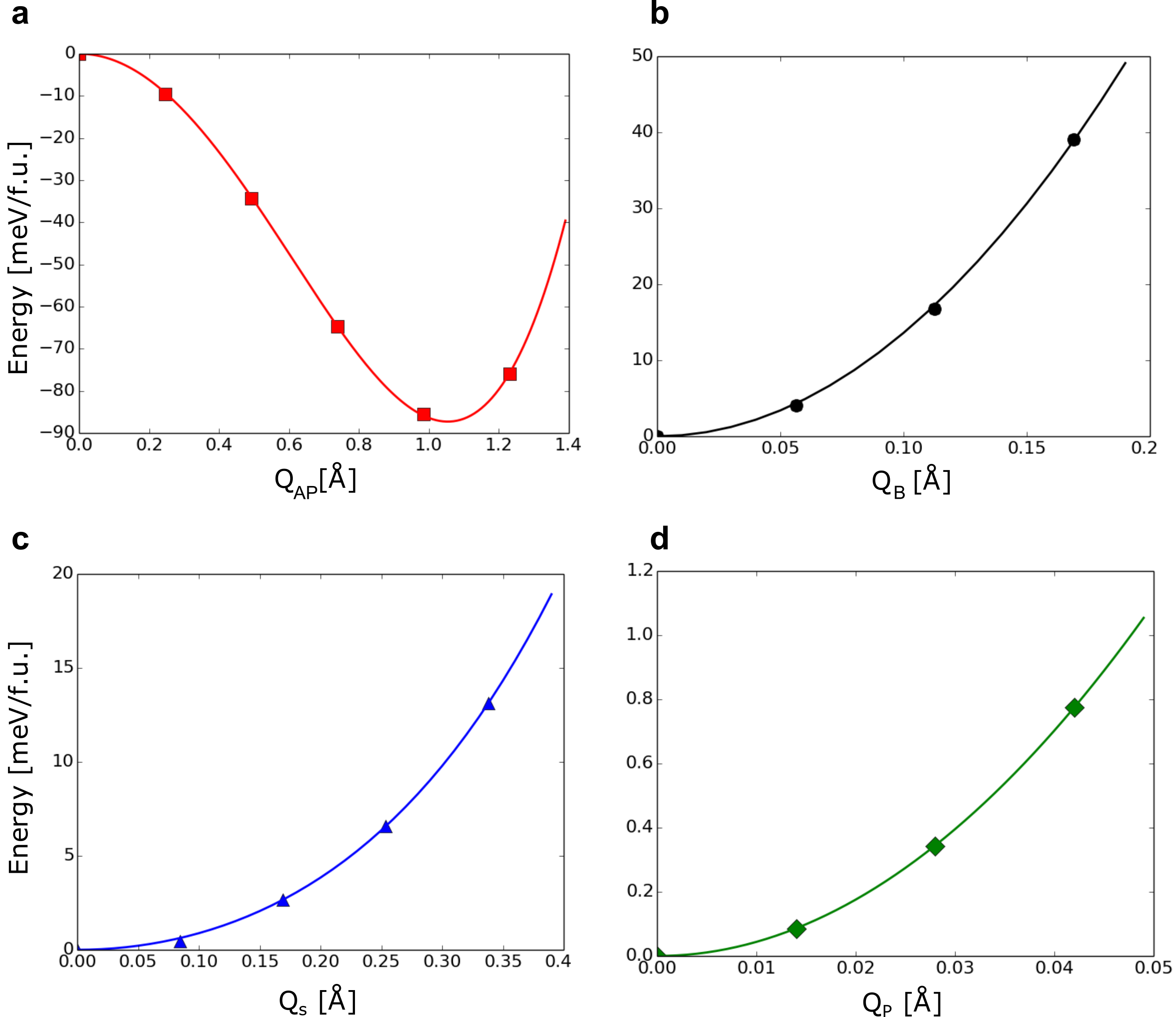}
\caption{\label{fig:surf1}Energy surfaces calculated with DFT by freezing in \textbf{a}, the $M_5^-$ distortion ($Q_{AP}$), \textbf{b}, the $M_4^+$ breathing distortion ($Q_B$), \textbf{c}, the $\Sigma_2$ distortion ($X1$-like displacements), and \textbf{d}, the $\Gamma_5^-$ ($Q_P$) polar distortion. By fitting these energy surfaces, we obtain the coefficients in equation (\ref{e:fulllandau}). The calculations are done in a 40 atom unit cell with A-AFM ordered along the $\mathbf{b}$-axis.}
\end{center}
\end{figure*}

\clearpage

\begin{figure*}
\begin{center}
\includegraphics[width=0.95\textwidth]{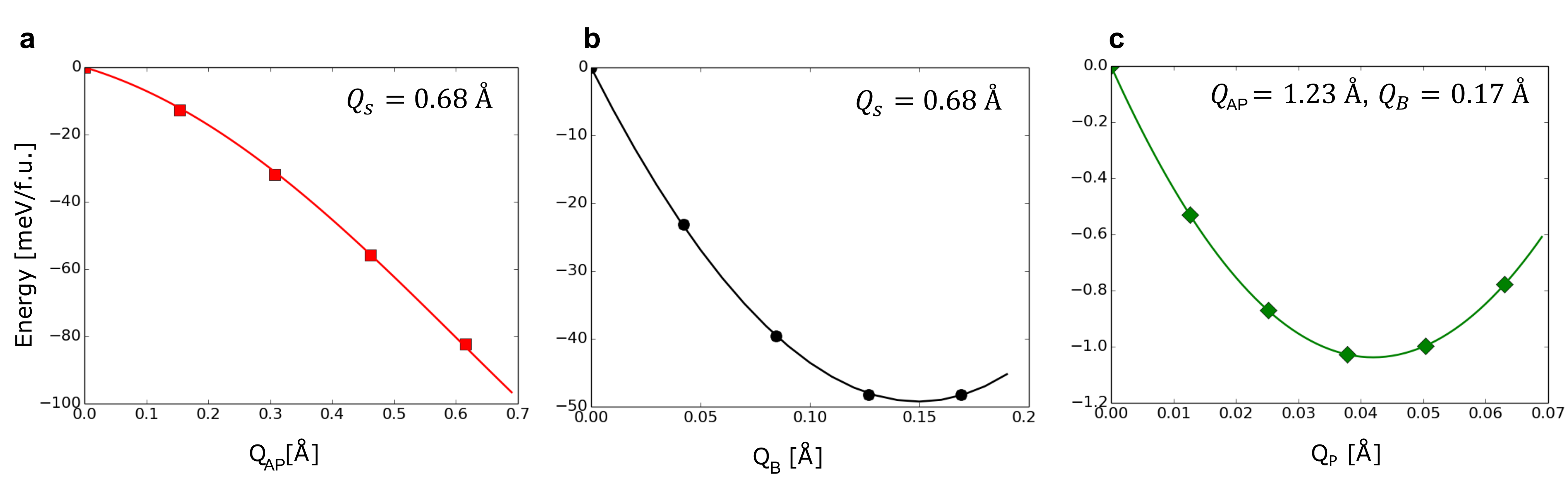}
\caption{\label{fig:surf2} Energy surfaces used to calculate the coupling coefficients in equation~(\ref{e:coupl}). 
\textbf{a,b}, The $\Sigma_2$ (X1-like) distortion is set to a fixed amplitude and then $Q_{AP}$, $Q_B$ is frozen in. 
\textbf{c}, Both $Q_{AP}$ and $Q_B$ are fixed, and the polar mode $Q_P$ is frozen in.}
\end{center}
\end{figure*}

\begin{figure*}
  \centering
  \includegraphics[width=0.999\textwidth]{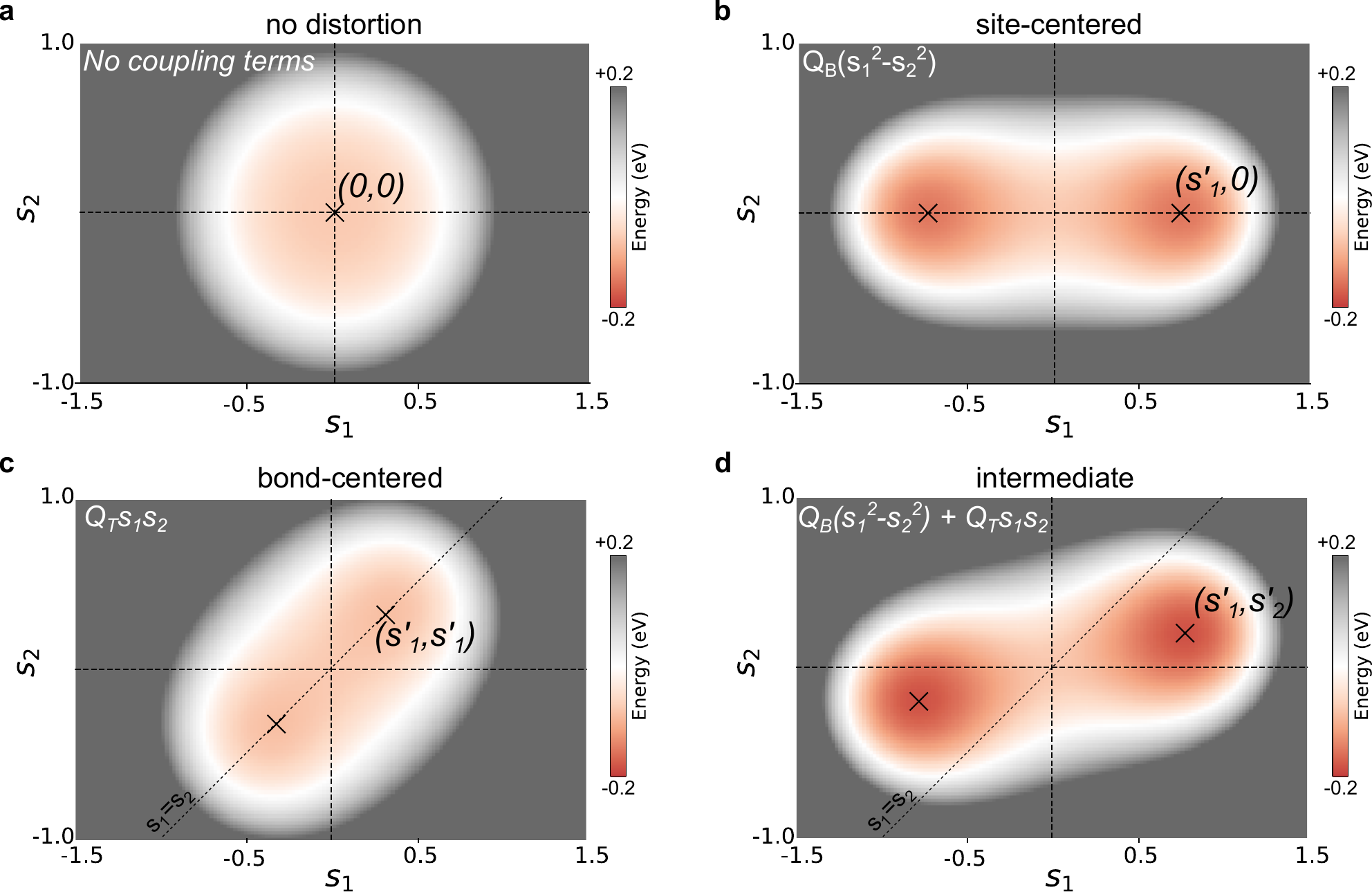}
  \caption{
        Energy surfaces based on the Landau theory.
\textbf{a}, If the third-order coupling terms are removed from equation~(\ref{e:fulllandau}), the displacements associated with the charge-ordered phase do not lower the free-energy.
\textbf{b}, Adding the $\delta_{bss}Q_{B}(s_1^2-s_2^2)$ coupling term lowers the energy substantially by having a finite amplitude in the X1-like displacement order parameter.
The energy is minimized (black cross) by having the X1-like amplitude equal to ($s'_{1}$,$0$) (site-centered order).
\textbf{c}, By including only the $\delta_{tss}Q_{AP}s_{1}s_{2}$ coupling term, the energy is minimized at the ($s'_{1}$,$s'_{2}=s'_{1}$) coordinate of the X1-like order parameter (bond-centered order). 
\textbf{d}, When both coupling terms are present, the energy is minimized at the ($s'_{1},s'_{2}\neq s'_{1}$) coordinate of the X1-like order parameter (intermediate order).
We set the amplitudes of $Q_{B}$, $Q_{AP}$, and $Q_{P}$ to the values in Supplementary Table~\ref{tab:decomp001} and the coefficients to those in Supplementary Table~\ref{tab:coeff}.
The picture does not change qualitatively by varying these values as long as the sign of the coefficients remains the same. 
The linecuts for Fig. 4 in the main text were taken from panels b and d, along the (0,$s_{2}$) directions and passing though the minimum energy positions (crosses).
}
  \label{F:Landau}
\end{figure*}